\documentclass[epj]{svjour}
\usepackage[usenames]{color}
\usepackage{cancel}
\usepackage{graphics}
\usepackage{amssymb}
\usepackage{threeparttable}  
\usepackage{lineno}
\begin{document}
\title{Periodic frequencies of the cycles in $2\times2$ games: evidence from experimental economics}
\author{Bin Xu\inst{1,2,3} \and Shuang Wang\inst{1,4,5}\and Zhijian Wang\inst{1,3}
\thanks{\emph{Present address:} }%
}                     
\offprints{}          
\institute{ Experimental Social Science Laboratory, Zhejiang University, Hangzhou, 310058, China
 \and Public Administration College, Zhejiang Gongshang University, Hangzhou, 310018, China
 \and  State Key Laboratory of Theoretical Physics, Institute of Theoretical Physics, Chinese Academy of Sciences, Beijing, 100190, China
 \and Chu Kochen College, Zhejiang University, Hangzhou, 310058, China
 \and Department of Economics, Duke University, Durham, NC, 27708, U.S.\\
 \email {wangzj@zju.edu.cn}}
\date{Received: date / Revised version: date}
%
\abstract{Evolutionary dynamics provides an iconic relationship---the periodic frequency of a game is determined by the payoff matrix of the game. This paper reports the first experimental evidence to demonstrate this relationship. Evidence  comes from two populations randomly-matched $2\times2$ games with 12 different payoff matrix parameters. The directions, frequencies and changes in the radius of the cycles are measured definitively. The main finding is that the observed periodic frequencies of the persistent cycles are significantly different in games with different parameters. Two replicator dynamics, standard and adjusted, are employed as predictors for the periodic frequency. Interestingly, both of the models could infer the difference of the observed frequencies well. The experimental frequencies linearly, positively and significantly relate to the theoretical frequencies, but the adjusted model performs slightly better.
} 
\maketitle
\section{Introduction}

Game theory, including classical game theory~\cite{VonNeumann1944} and evolutionary game theory~\cite{Maynard1982evolution}, provides a mathematical framework for systems ranging from microbial~\cite{Frey2010} to social economy~\cite{myerson1997game}. However, the theoretical predictions of classical game theory and evolutionary game theory sometimes differ, two typical examples of this being the  Matching Pennies game~\cite{Dawkins1976,Sigmund1981,cremer2008anomalous,claussen2007drift} and the Rock-Paper-Scissors game. In these games, evolutionary game theory predicts clear cyclic motions in social state space~\cite{Taylor1978,Sigmund1981,Weibull1997,Hofbauer2003,Sandholm2011}, while classical game theory does not.

 As a major source of knowledge in social sciences~\cite{Falk2009}, laboratory experiments have been employed to test evolutionary game theory~\cite{Samuelson2002}. Although experiments have supported the existence of the cycles~\cite{Samuelson2002,Binmore2001,Friedman1996,Plott2004globalScarf}, the definitive measurement of cycles has rarely been reported. Quite recently, in a continuous time and centralised information protocol, cycles have been found in Rock-Paper-Scissors (RPS)-like games~\cite{Friedman2012}. In addition, Xu, Zhou and Wang have studied cycles in standard RPS games with a random matching and decentralised information protocol~\cite{Xu20134997}. They proved that cycles do exist in continuous time RPS-like games~\cite{Friedman2012} and discrete time standard RPS games~\cite{Xu20134997}. Additionally, the directions of all of the cycles detected in these RPS experiments met the evolutionary dynamics very well~\cite{Friedman2012,Xu20134997}. Therefore, we can say that the cycles in the experiments are \emph{qualitatively} consistent with the expectations of evolutionary game theory.

Subsequently, a straightforward question arises, namely what is the quantitative relationship between cycles in experiments and the parameters of the games? To answer this question, we look to the central frequencies of the cycles which can be (1) measured definitively in experiments and (2) explicitly calculated based on dynamics equations.

Our data is from Selten and Chmura's experiments of twelve $2\times 2$ games~\cite{selten2008} (see Fig.~\ref{fig:seltenpayoff} and Table~\ref{payoffmatrix12andFreq}). According to replicator dynamics models~\cite{Taylor1978,Maynard1982evolution,Weibull1997,Traulsen2005}, the payoff matrices of the twelve games indicate that the transits among social states in each game should form endless cycles. As mentioned in Ref.~\cite{Nowak2012}, evolutionary game theory has rarely been supported by human experiments. In this article, we will reveal the cycles in the 12 experiments.

With methods rooted in non-equilibrium statistical physics~\cite{Xu20134997}, we counted the cycles chronologically in the experiments. The main results include: (1) cycles exist; (2) cycles are persistent; (3) the empirical cyclic frequencies in the 12 games are quantitatively different, see Table~\ref{tab:theta}. Moreover, the differences could be well described by the theoretical frequencies derived from replicator dynamics models, see Table~\ref{tab:OLRresults}.

The structure of this paper is as follows. In the next section, we will briefly introduce the 12 experiments and provide the theoretical predictions for the cycles. In Section 3, the measuring methods for the cycles are explained. Section 4 summarises the results. The last section includes discussions and summary.

\section{Experiments and Theoretical expectations}

To capture the cyclic social behaviours in laboratory game experiments, we use the existing well-known experimental data of Selten and Chmura~\cite{selten2008} and refer to two theoretical models of evolutionary dynamics~\cite{Taylor1978,Maynard1982evolution,Weibull1997,Traulsen2005}, standard replicator dynamics and adjusted replicator dynamics.

\subsection{Two-population $2\times2$ games experiments}\label{sectionExperimentSelten}

The original aim of Setlen and Chumra's experimental investigation is to compare the performance of the five stationary concepts~\cite{selten2008}. They designed twelve $2\times2$ cyclic matching-pennies-like games~\cite{Traulsen2005,cremer2008anomalous}. Games 1 to 6 are constant-sum games, and each of them contains 12 groups; games 7 to 12 are non-constant-sum games, and each of them contains 6 groups. In addition, games 7 to 12 have the same Nash equilibria as the games 1 to 6, respectively.

\begin{figure}
\centering
\resizebox{0.15\textwidth}{!}{
\includegraphics{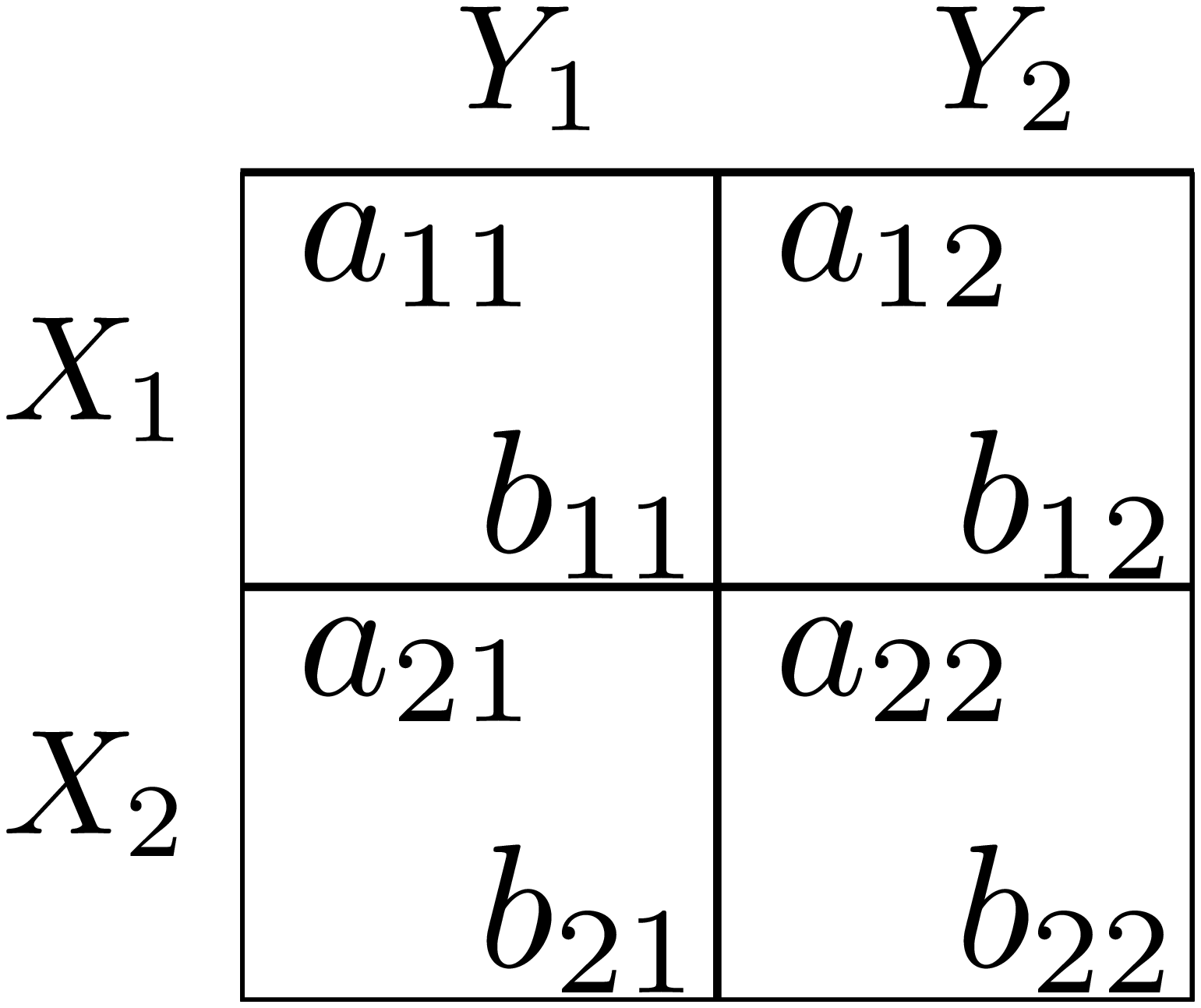}
}~~~~
\caption{Payoff matrix of the $2\times2$ game. For the 12 games, the elements of the payoff matrix  are different,  see Table~\ref{payoffmatrix12andFreq}.
\label{fig:seltenpayoff}}
\end{figure}

Each of the 108 groups consists of two four-person populations, so a total of  864 subjects are involved in the experiments. Each group plays the game for 200 rounds repeatedly. The strategy set for each player in the first population (denoted by $X$) is ($X_{1}$,$X_{2}$); while the strategy set for each player in the second  population (denoted by $Y$) is ($Y_{1}$,$Y_{2}$). In each round, every subject plays the game against an opponent randomly picked from the other population. The payoff matrix for each game has the same form as Fig.~\ref{fig:seltenpayoff} shows, where $a_{ij}$ ($b_{ij}$) indicates the payoff for the player in the first (second) population, $i,j=1,2$. The specific value of $a_{ij}$ ($b_{ij}$) in each game is exhibited in Table~\ref{payoffmatrix12andFreq}.

\subsection{Predictions from two replicator dynamics}

As an approximation, evolutionary game dynamics can be a powerful tool to describe finite population games \cite{Borgers1997,Friedman1998,Samuelson2002,Friedman2012}. It takes the social state, i.e. the combination of the densities of strategies in each population~\cite{Sandholm2011}, as a crucial object and analyses its motion in state space. In this case, each social state is denoted as $x$:=$(p, q)$, where $p$~($q$) is the density of $X_1$~($Y_1$) in the first (second) population $X(Y)$ and the state space is $\mathbb{X}$=$\{0,\frac{1}{4},\frac{2}{4},\frac{3}{4},1\} \times\{0,\frac{1}{4},\frac{2}{4},\frac{3}{4},1\}$, as in Fig.~\ref{fig:5x5lattice} (for more details, see Appendix A). According to evolutionary game dynamics, the social state will change in the state space.

As predictors, the two models---standard replicator dynamics and adjusted replicator dynamics---are used as references. The advantage of the two models is that they are parameter free and the most widely used~\cite{Traulsen2005,Maynard1982evolution,Sandholm2011,Friedman1996,cremer2008anomalous}. The solutions in a (sufficiently small) neighbourhood of the Nash equilibrium can be calculated for the two dynamics for the twelve games. According to the calculation, the motions in these games are cyclic (see Appendix~\ref{sectionTheoFreq}) and equivalent to two-dimensional \emph{simple harmonic motions} (see Appendix J in~\cite{Maynard1982evolution}). Therefore, based on these two models, the observed social state is supposed to keep moving in the state space~\cite{Taylor1978,Maynard1982evolution,Weibull1997,Traulsen2005} and thus form endless cycles, which leads to the following theoretical predictions:
\begin{enumerate}
  \item The cycles, for all 12 games, exist and persist. Meanwhile, the directions of the cycles are all clockwise.
  \item The frequencies of the cycles are different (for numerical results see Table~\ref{payoffmatrix12andFreq}, in which $f_s$ and $f_a$ respectively indicate the predictions from standard and adjusted replicator dynamics).
  \item The radius of the cycles, for all 12 games, should remain unchanged over time.
\end{enumerate}

With 12 experimental $2\times2$ games, we verify the existence and persistence of the cycles; and test the frequencies of cycles quantitatively. Furthermore, we find that the observed periodic frequencies of the cycles are different among games. Moreover, the experimental frequencies are linearly, positively and significantly related to the theoretical frequencies.

 \begin{table*}
\centering
\renewcommand{\arraystretch}{1.3}
\caption{Payoff matrices, equilibrium, eigenvalues and theoretical frequencies.}\label{payoffmatrix12andFreq}
\begin{tabular}{||c||rr|rr|rr|rr||rr||cc||cc||}
    \hline
Game	&$a_{11}$	&$b_{11}$	&$a_{12}$	&$b_{12}$	&$a_{21}$	&$b_{21}$	&$a_{22}$	 &$b_{22}$& ${x_{N}}^\S$	 & $y_{N}$  &$\lambda^{2}_s$$^\dag$	 &$\lambda^{2}_a$&$f_{s}$$^\ddag$ &$f_{a}$\\
   \hline
1	&	10	&	8	&	0	&	18	&	9	&	9	&	10	&	8	&	0.91&	0.09&	 -0.83&	-0.01&	 0.1447&	 0.0161\\
2	&	9	&	4	&	0	&	13	&	6	&	7	&	8	&	5	&	0.73&	0.18&	 -3.57&	-0.08&	 0.3007&	 0.0463\\
3	&	8	&	6	&	0	&	14	&	7	&	7	&	10	&	4	&	0.91&	0.27&	 -1.98&	-0.04&	 0.2241&	 0.0320\\
4	&	7	&	4	&	0	&	11	&	5	&	6	&	9	&	2	&	0.82&	0.36&	 -4.17&	-0.14&	 0.3248&	 0.0591\\
5	&	7	&	2	&	0	&	9	&	4	&	5	&	8	&	1	&	0.73&	0.36&	 -5.55&	-0.28&	 0.3751&	 0.0841\\
6	&	7	&	1	&	1	&	7	&	3	&	5	&	8	&	0	&	0.64&	0.46&	 -6.94&	-0.45&	 0.4193&	 0.1071\\
7	&	10	&	12	&	4	&	22	&	9	&	9	&	14	&	8	&	0.91&	0.09&	 -0.83&	-0.01&	 0.1447&	 0.0155\\
8	&	9	&	7	&	3	&	16	&	6	&	7	&	11	&	5	&	0.73&	0.18&	 -3.57&	-0.07&	 0.3007&	 0.0419\\
9	&	8	&	9	&	3	&	17	&	7	&	7	&	13	&	4	&	0.91&	0.27&	 -1.98&	-0.03&	 0.2241&	 0.0297\\
10	&	7	&	6	&	2	&	13	&	5	&	6	&	11	&	2	&	0.82&	0.36&	 -4.17&	-0.11&	 0.3248&	 0.0537\\
11	&	7	&	4	&	2	&	11	&	4	&	5	&	10	&	1	&	0.73&	0.36&	 -5.55&	-0.21&	 0.3751&	 0.0734\\
12	&	7	&	3	&	3	&	9	&	3	&	5	&	10	&	0	&	0.64&	0.46&	 -6.94&	-0.31&	 0.4193&	 0.0880\\
\hline
\end{tabular}
\begin{flushleft}
$^\S$ $x_N(y_N)$ is the Nash equilibrium for the X (Y) population \cite{selten2008}; $^\dag$ $\lambda^{2}_s$ ($\lambda^{2}_a$) is the square of the eigenvalue calculated from standard (adjusted) replicator dynamics Eq.~\ref{eq:lambda2SRD} (Eq.~\ref{eq:lambda2ARD}) in Section~\ref{eigenvalues}; $^\ddag$ $f_{s}$ ($f_{a}$) is the theoretical frequency of the standard (adjusted) replicator dynamics. \\
\end{flushleft}
\end{table*}

\section{Measurement}\label{secMethods}

In evolutionary game theory, social state is used to describe social evolution~\cite{Sandholm2011}. For the 12 experiments of finite population in this article, the state space is discrete~\cite{XuWang2011ICCS,Nowak2012,Friedman2012}. As illustrated in Fig.~\ref{fig:5x5lattice}, there are altogether 25 social states (the explanations are given in~\ref{25states} in the Appendix). For a certain round $t$, the social state must be one of the 25 states. Across the experimental rounds, the observed state should transit, and a stochastic-like trajectory should be formed~\cite{Sandholm2011,Friedman2012,Friedman1998,Samuelson2002}.

In stochastic dynamics processes, as emphasised by Young~\cite{Young2008}, in the detailed balance condition, the transitions between any pair of  states is symmetric. According to classical game theory, no net transit can be found in the state space and the system is in equilibrium under this condition.

On the contrary, evolutionary game theory predicts that the transits among the states are asymmetrical. For example, in Fig~\ref{fig:5x5lattice}, transit 3$\rightarrow$4 should be more frequent than the reverse (4$\rightarrow$3), because transit 3$\rightarrow$4 corresponds to the direction of the cycles moving around the Nash equilibrium. Therefore, the detailed balance symmetry should be broken and a pattern of net transit (cycles) must emerge in the long run~\cite{XuWang2011ICCS,kinkPleimling2011E,XWMm2011}.

\subsection{Cycle counting}

\begin{figure}
\centering
\resizebox{0.35\textwidth}{!}{
\includegraphics{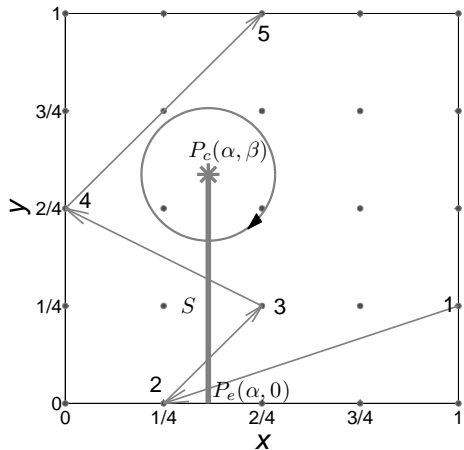}
}~~~~
\caption{Schematic diagram for counting cycles~\cite{Friedman2012,Xu20134997}. Background lattice represents the 25 social states. Including 4 transits, the successive segments (1 $\rightarrow$ 2 $\ldots$ $\rightarrow$ 5) form a trajectory of a social evolutionary transit path. The segment $S$ is the Poincare section (tripwire) from $P_{c}:=(\alpha, \beta)$ (star) to $P_{e}:=(\alpha, 0)$ at the lower edge. In this paper, for each game, we set $(\alpha, \beta)$ at its Nash equilibrium ($x_N, y_N$). For example, for Game-1, $(\alpha, \beta) =(0.91,0.09)$. The direction of the cycle surrounding the centre $(\alpha, \beta)$ is clockwise, as predicted by the two models.
\label{fig:5x5lattice}}
\end{figure}

To verify the cycles quantitatively, we look to Poincare section (tripwire)~\cite{Friedman2012,Xu20134997} for counting cycles. First, we set a segment $S$ called the Poincare section~\cite{Friedman2012}, as shown in Fig.~\ref{fig:5x5lattice}, from $P_{c}=(\alpha, \beta)$ to $P_{e}=(\alpha, 0)$, and we take $(\alpha, \beta)$ equal to the Nash equilibrium point ($x_{N},y_{N}$) (see Table~\ref{payoffmatrix12andFreq}) of each game. A transit is a directional segment from state $(x_{1}, y_{1}$) observed at time $t$ to state $(x_{2}, y_{2})$ at time $t+1$. These two segments could intersect at X as
\begin{equation}\label{Xcross}
    X:\equiv(X_{x},X_{y})=\left(\alpha,y_{1}+\frac{(y_{2}-y_{1})(\alpha-x_{1})}{x_{2}-x_{1}}\right)
\end{equation}
According to Ref~\cite{Xu20134997}, the value of $C_{i}$ for a transit from time $t$ to $t+1$ should be:
\begin{enumerate}\small
  \item $C_{i}=0$, if $X\not\in(P_{c},P_{e}]\cup x_{2}=x_{1}$
  \item $C_{i}=1$, if $X \in(P_{c},P_{e}]\cap x_{2}>\alpha>x_{1}$
  \item $C_{i}=-1$, if $X \in(P_{c},P_{e}]\cap x_{2}<\alpha<x_{1}$
  \item $C_{i}=\frac{1}{2}$, if $X \in(P_{c},P_{e}]\cap x_{2}>x_{1}\cap(x_{1}=\alpha\cup x_{2}=\alpha)$
  \item $C_{i}=-\frac{1}{2}$, if $X \in(P_{c},P_{e}]\cap x_{2}<x_{1}\cap(x_{1}=\alpha\cup x_{2}=\alpha)$
\end{enumerate}
The accumulated counting number $C_{t_{0}},_{t_{1}}$ of the evolutionary trajectory during the time interval [$t_{0}, t_{1}$] is defined ( see Eq.2 in~\cite{Xu20134997}) as
\begin{equation}\label{eqcri}
    C \equiv \sum_i  C_i.
\end{equation}
The accumulated counting number $C$ quantifies the net number of cycles around the reference point ($\alpha, \beta$). As mentioned~\cite{Xu20134997}, such a quantity can help us to detect deterministic behaviours in a stochastic process. {$C$} indicates the existence of cycles when it does not equal 0. Additionally, $C>0 $ ($C<0$) indicates that the direction of the cycles is counterclockwise (clockwise). We use the subscript $(t_0,t_1)$ to specify the observation in time interval $[t_0,t_1]$, like $C_{t_0,t_1}$. Clearly, $C$ is time odd ($C_{t_0,t_1}$ = - $C_{t_1,t_0}$). By calculating $C$ and investigating how it  changes with time, we can verify the existence, persistence and direction of the cycles.

\subsection{Measurement of the frequency  of the cycles}\label{secMeasureFreq}

In this article, the experimental periodic frequency ($f_{exp}$) is defined as the time averaged net cycles counted in an experimental trajectory, which referring to~\cite{Xu20134997}, is
\begin{equation}\label{eq:expfrequency}
 f_{exp} =\frac{\sum_i C_i}{t_1 - t_0},
\end{equation}
where $\sum_i C_i$ is the accumulated counting number of cycles in time interval [$t_0,t_1$] (the time unit in these 12 experiments is discrete in the experimental round), and $\bar{f}_{exp}$ is the average ${f}_{exp}$ over 12 groups of each of games 1 to 6 or 6 groups of each of games 7 to 12.

\subsection{Measurement of the radius  of the cycles}\label{secMeasureradius}

Whenever a  transit-$i$ crosses the Poincare section, the Euclidean distance (denoted as $x_i$) between the crossing dot and the centre $N$ can be measured. Furthermore, for a given time interval $[t_0,t_1]$, the average radius of the cycles\footnote{This observable, as a vector, can be understood as the moment of force or torque in classical mechanics in physics.} can be defined as
\begin{equation}\label{eqR}
    R_{t_0,t_1} = \frac{\sum_i x_i C_i}{\sum_i |C_i|}
\end{equation}
where $i$ (=$1,2,3,\cdots$) denotes the $i$-th time that the trajectory passes the Poincare section. The sum covers all of the transits across the Poincare section. For example, according to Eq.~\ref{eqR}, the radius of the net cycles ($R_{1,5}$) in Fig.~\ref{fig:5x5lattice} is -0.115. We use this measurement to evaluate the change in radius with time, which is necessary for cycle testing~\cite{Nowak2012,Friedman2012}.

\section{Results}

As mentioned above (see Fig.~\ref{fig:5x5lattice} and the related text), we can illustrate the evolutionary trajectories of each group in the 12 experiments with the experimental data (see section \ref{sectionExperimentSelten}). Accordingly, using the methods described in section~\ref{secMethods}, we can measure the experimental observations (accumulated number, frequency and radius of cycles) of the 12 experiments. Then, the experimental observations can be compared with the theoretical expectations (see section~\ref{secTheoVal}) of the two different models. Results of statistical analysis are reported as follows.

\subsection{Existence, persistence and direction of the cycles}

\textbf{Result 1:} Cycles exist and persist, and the directions of net cycles in the ``stochastic" trajectories meet the predictions of both evolutionary dynamics.

To test the existence of net cycles as in~\cite{Xu20134997}, we calculate the average value $\bar{C}_{1,200}$ for the 12 games as well as the standard error of the mean (s.e.m) over the groups. The results are shown in Table \ref{tab:theta}. Using the Wilcoxon signed-rank test, we find  that the $\bar{C}_{1,200}$ of all games are significantly less than 0, indicating the existence of cycles in all games  except game 9,\footnote{However, two points should be mentioned. First, each of games 7 to 12 only has 6 groups while each of games 1 to 6 has 12 groups; second, theoretically, the frequencies of games 1, 7, 3 and 9 are small, indicating that the formation of each cycle needs more time. So we use half of the observations of games 1 to 6, again, $p$-value$>5\%$ in games 1, 2, 3 and 4. So we still use this empirical number of cycles when comparing with the theoretical value later.} see Table~\ref{tab:theta}. Using the 12 games as samples, we find that the existence of the net cycles is significant ($p$=$0.002$, Wilcoxon signed-rank test with $C_{1,200}=0$).

\begin{table*}[htbp2]
\caption{Accumulated C, frequency and radius of the net cycles in the 12 experiments.}\label{tab:theta}
\centering
\renewcommand{\arraystretch}{1.3}
 \small
\begin{tabular}{|c|r|rrr|rr|rrrr|}
   \hline
Game	&Obs.$^\flat$ &$\bar{C}_{1,200}$	&$s.e.m^\dag$	&$p$-value$^\ddag$	 &$\bar{f}_{exp}$	 &$s.e.m^\dag$	 &$\bar{R}_{1,100}$	 &$\bar{R}_{101,200}$	 &$\bar{D}^\S$	&$p$-value$^\ddag$\\
  \hline
  1&	12&	-3.167&	0.976&	0.003&	0.016&	0.005&	-0.035&	-0.018&	0.017&	0.166\\
2&	12&	-4.000&	1.142&	0.013&	0.020&	0.006&	-0.017&	-0.026&	-0.009&	0.184\\
3&	12&	-2.750&	0.789&	0.006&	0.014&	0.004&	-0.016&	-0.010&	0.006&	0.386\\
4&	12&	-5.917&	1.062&	0.003&	0.030&	0.005&	-0.017&	-0.015&	0.002&	0.488\\
5&	11&	-8.167&	1.160&	0.002&	0.041&	0.006&	-0.022&	-0.024&	-0.002&	0.751\\
6&	12&	-10.083&	1.311&	0.002&	0.051&	0.007&	-0.026&	-0.020&	0.006&	0.204\\
7&	6&	-1.333&	0.211&	0.023&	0.007&	0.001&	-0.014&	0.000&	0.014&	0.423\\
8&	6&	-4.667&	1.606&	0.046&	0.023&	0.008&	-0.017&	-0.025&	-0.008&	0.423\\
9&	6&	-2.333&	1.282&	0.091&	0.012&	0.006&	-0.010&	-0.006&	0.004&	0.873\\
10&	6&	-6.333&	1.745&	0.027&	0.032&	0.009&	-0.016&	-0.016&	0.000&	0.522\\
11&	6&	-12.000&	1.125&	0.027&	0.060&	0.006&	-0.024&	-0.026&	-0.001&	0.749\\
12&	6&	-9.000&	2.408&	0.027&	0.045&	0.012&	-0.021&	-0.017&	0.004&	0.631\\
  \hline
\end{tabular}
\begin{flushleft}
$^\flat$: Number of groups. $^\dag$: Standard error of the mean. $^\ddag$: Wilcoxon signed-rank test, two-tailed. $^\S$: Average difference between $\bar{R}_{1,100}$ and $\bar{R}_{101,200}$ over 12 (6) groups.\\
\end{flushleft}
\end{table*}

The Wilcoxon signed-rank test verifies the persistence of cycles by comparing $C_{1,100}$ and $C_{101,200}$ of the 12 games. Results show that the null hypothesis: $C_{1,100}$=$C_{101,200}$ cannot be rejected ($p$=0.695, two-tailed, $d.f.$=11), which indicates the persistence of cycles.

Since all $\bar{C}_{1,200}$ of the 12 games are negative (see Table \ref{tab:theta}), the empirical directions of the cycles are clockwise statistically, consistent with the directions predicted by both of the two dynamics models as shown in Fig.~\ref{fig:12DynaSelten}.

\subsection{Frequencies of the cycles}\label{sectionFreq} 
\textbf{Result 2:} The periodic frequency of the net cycles can be tested in data.

\begin{figure*}
\centering
\resizebox{0.6\textwidth}{!}{
\includegraphics{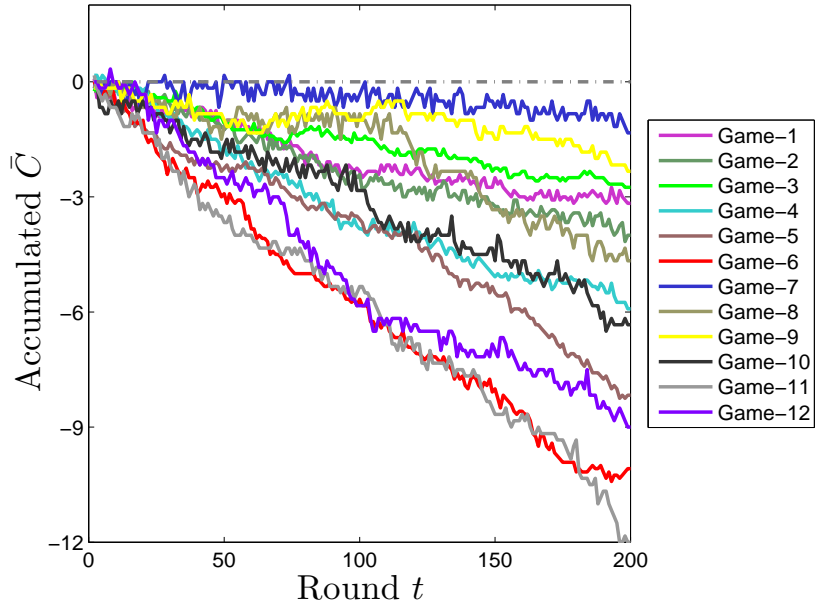}}
\caption{The dependence of the counted cycles on time. Each line is the average accumulated $C$ for one game. Negative values of the average accumulated  $C_{1,t}$ indicate that the cycles are clockwise. Successive decline of the values indicates the persistence of cycles along time. The red curve reports the results of Game-6, in which the accumulated $\bar{C}$ is almost -10, indicating 10 clockwise cycles were obtained in 200 rounds and the frequency of cycles is about 0.05 (per round).
\label{fig:RotationSelten}}
\end{figure*}

Before calculating the frequencies of cycles as in~\cite{Xu20134997}, the number  of cycles in the trajectory is counted. Fig.~\ref{fig:RotationSelten} presents the average number of the net cycles ($C_{1,t}$, see Eq.~\ref{eqcri} for its explanation) along time $t$ ($t$ is up to 200 rounds in experiments, $t=200$) of each game. The 12 curves respectively represent the average cycles counted for the 12 experiments. For example, the red curve (Game 6) ends at about -10 at $t=200$. This indicates that in the 200-round game, about 10 cycles are counted on average over the 12 groups; meanwhile, the negative value implies the clockwise direction of the cycles.

The frequencies $f_{exp}$ of the 12 experiments are calculated according to Eq.~\ref{eq:expfrequency}. The results are shown in Table~\ref{tab:theta}. The observed periodic frequencies of the 12 games are significantly different, which is consistent with the patterns of Fig.~\ref{fig:RotationSelten}. Specifically, Eq.~\ref{eq:expfrequency} indicates that the frequencies can be represented by the slopes of the $C_{1,200}$ curves in Fig.~\ref{fig:RotationSelten}. Since the slopes are different for the 12 games, the frequencies are different.

In Section~\ref{ModelTestDisc}, we will show that the empirical frequencies of the cycles is an appropriate criterion for evaluating the models quantitatively.

\subsection{Change of the cycle radius over time}

\textbf{Result 3:} The radius of the cycles does not change with time in the experiments, which supports the predictions of both of the two models.

For the 200-round experiments, our hypothesis is that the radius of the first 100 rounds ($R_{1,100}$) is equal to that of the second 100 rounds ($R_{101,200}$). The calculated results of $R_{1,100}$ and $R_{101,200}$ of the 12 games are shown in Table~\ref{tab:theta}. Using the groups in each game as samples, the radius does not change in each of the 12 games ($p$-values are exhibited in Table~\ref{tab:theta}, last column), and therefore the equal hypothesis cannot be significantly rejected. Using the 12 games as samples, we find that the equal hypothesis cannot be significantly rejected ($p=0.308$, Wilcoxon signed-rank test). Therefore, we can say that the statement that the radius does not change with respect to  time is statistically supported.

\section{Frequency as a criterion for model testing}\label{ModelTestDisc}

As mentioned above, as a novel index, empirical frequencies of cycles could be employed to evaluate theoretical models quantitatively. We regress theoretical frequency ($f_{s}$ and $f_{a}$ in Table~\ref{payoffmatrix12andFreq}) on experimental frequencies ($\bar{f}_{exp}$ in Table~\ref{tab:theta}) with the method of OLR and display the results in Table~\ref{tab:OLRresults} (Appendix~\ref{ModelEval} provides details). These statistical results state that both of the models can interpret the differences among the observed frequencies quite well ($t>5$ in Table~\ref{tab:OLRresults}), while the adjusted model ($t$-value 6.41) performs slightly better.

In game theory, there exist tens of models known as population dynamics models (e.g.,~\cite{Weibull1997,Hofbauer2003,Sandholm2011,Castellano2009,szabo2007evolutionary}), finite population models (e.g.,~\cite{alos2003finite,Traulsen2005,cremer2008anomalous}) or learning models (e.g., ~\cite{Camerer2003,RothComp2010}). Comparisons between the models can be important~\cite{Friedman2012,Xu20134997}, however since the main purpose of this paper is to report the definitive measurements in experiments, we do not extend our theoretical analysis to such comparisons. Nevertheless, the clear result ($t>5$ in Table~\ref{tab:OLRresults}) indicates that dynamic game models could be used effectively in combination  with laboratory game experiments.

 \begin{table}[htbp2]
\centering
\small
\caption{\label{tab:OLRresults} OLR results of $\bar{f}_{exp}$ vs. $f_{s}$ and $f_{a}$ for the two models.}
\renewcommand{\arraystretch}{1.3}
\begin{tabular}{|c||rrrrr|}
  \hline
$\bar{f}_{exp}$	&$Coef.$	&$s.e.m$	&$t$~~	 &$P>|t|$	&[~95\%~~C.I.~]	 \\
   \hline
$f_{s}$    & 0.157     &0.026      &6.01        &0.000     &~0.099~~0.215\\
$\_cons$	    &-0.018     &0.008      &-2.16       &0.056     &-0.036~~0.001\\
   \hline
$f_{a}$	&0.519	    &0.081	    &6.41	     &0.000	    &$~0.338~~0.699$	\\
$\_cons$	    &0.001	    &0.005	    &0.25	     &0.805	    &$-0.010~~0.012$	\\
  \hline
\end{tabular}
\end{table}

\section{Discussion and summary}

First, in the 12 experimental $2\times2$ games, we have confirmed the existence and persistence of the cycles, and the frequencies of cycles have been tested quantitatively. Furthermore, we have found that the observed periodic frequencies of the cycles are different when the games differ. Moreover, the experimental frequencies are linearly, positively and significantly related to the theoretical frequencies.

Experimental confirmation of social behaviour  cycles is difficult~\cite{benaim2009learning}. Researchers continue to seek  significant cyclic patterns to support   evolutionary dynamics~\cite{Friedman2012,Nowak2012,XuWang2011ICCS,Xu20134997}. In some earlier works~\cite{wang2012evolutionary}, angular momentum was employed as an observable, but no periodic frequency was reported. In Ref.~\cite{Friedman2012}, CRI was developed to index cycles, but once again no information was provided about periodic frequency~\cite{Friedman2012}. In Ref.~\cite{Xu20134997}, frequency served as an index for cycles, howeverthere is only one payoff matrix in their experiments, so that a quantitative comparison  with theoretical frequencies is impossible. To the best of our knowledge,   this paper is the first to report that cycles' frequencies depend on the payoff matrices of the corresponding  games.

We wish to emphasise our puzzles as follows. Even though the theoretical frequencies ($f_s$ and $f_a$) could meet experimental frequencies ($\bar{f}_{exp}$) well by OLR (see Table~\ref{tab:OLRresults}), we still hope to emphasise that such \emph{correspondence} could be a coincidence, for example because the differences in the observed frequencies ($\bar{f}_{exp}$) could be interpreted by other factors, e.g., the distance ($D_{c}:=[(x_{N}-0.5)^2+(y_{N}-0.5)^2]^{\frac{1}{2}}$) from the centre of state space (0.5, 0.5) to the Nash Equilibrium ($x_{N},y_{N}$) of a game. For these 12 games, by OLR,  there is a significant dependence of $\bar{f}_{exp}$    on   $D_{c}$  ($t>5$) too, see Table~\ref{tab:OLRresults2}. We are unaware of the reason for this significance.  At the same time, we are unaware of why the coefficient in Table~\ref{tab:OLRresults} differs significantly from 1. Admittedly, the relationship between games' payoff matrix and empirical frequency is unclear. We hope to return to these issues in the future.

 \begin{table}[htbp2]
\centering
\small
\caption{\label{tab:OLRresults2}OLR results of $\bar{f}_{exp}$ and the distance between the Nash Equilibrium and the centre of state space.}
\renewcommand{\arraystretch}{1.3}
\begin{tabular}{|c||rrrrr|}
  \hline
$\bar{f}_{exp}$	&$Coef.$	&$s.e.m$	&$t$~~	 &$P>|t|$	&[~95\%~~C.I.~]	 \\
   \hline
$D_{c}$    & -0.104     &0.017      &-6.01        &0.000     &-0.142~~-0.065\\
$\_cons$	    &0.067     &0.007      &9.92       &0.000     &0.052~~~0.082\\
  \hline
\end{tabular}
\end{table}

In summary, we report the existence of persistent cycles, as well as their directions in $2\times2$ games. Furthermore, the numerical frequencies of cycles are measured definitively and found to be significantly different across the 12 games (see $\bar{f}_{exp}$ in Table~\ref{tab:theta}). We believe these empirical frequencies of cycles can be interpreted more precisely.\\

\textbf{Acknowledgments}: We, WZJ and XB, thank Hai-Jun Zhou for his critical comments and generous support  on this work. We thank the anonymous reviewers for the helpful suggestions and Dr. Ekka for the excellent proofreading. We thank also Chuang Wang's suggestions and Zunfeng Wang's technical assistance. This work was supported by grants from the 985 Project at Zhejiang University (Project for experimental social science), SKLTP of ITPCAS (No. Y3KF261CJ1) and Philosophy $\&$ Social Sciences Planning Project of Zhejiang Province (13ND-JC095YB). Author Contributions: WZJ conceived research; WZJ and XB performed experimental analysis; WZJ and WS  performed theoretical analysis;  XB, WS and WZJ wrote the paper. After moved to Duke,  WS did the language revision.

\newpage
\large
\appendix{Appendix}

\section{Theoretical predictions on cycles}\label{sectionTheoFreq}

Two versions of replicator dynamics, standard replicator dynamics and adjusted replicator dynamics, are used as references in the 12 experiments. As mentioned in main text,  these   parameter-free models are the most widely applied in both the economics and physics literature~\cite{Traulsen2005,Maynard1982evolution,Sandholm2011,Friedman1996,cremer2008anomalous,Weibull1995evolutionary}.
Using the standard methods (e.g.~\cite{Weibull1997,Sigmund1981}), we can replicate the theoretical results on cycles.

\subsection{Social states}\label{25states}

In the view of evolutionary game theory~\cite{Sandholm2011}, at each moment $t$, the consequence of social interaction can be described as a social state, which is a vector with the densities of strategies in each population as components. In the two-population game here, the strategy set for each player in the row population is (up, down) or $(X_{1}$,$X_{2})$ while for each player in the column population is (left, right) or $(Y_{1}$,$Y_{2})$. Let $p$~($q$) denote the density of $X_1$~($Y_1$) in the row (column) population, then $x$:=$(p, q)\in \mathbb{X}$ represents a social state, herein  $\mathbb{X}$ is the social state space. Therefore, the state space is always a unit square as shown in Fig.~\ref{fig:5x5lattice}.

Because there are four subjects in each population in this game, the social state space is $\mathbb{X}$=$\{0,\frac{1}{4},\frac{2}{4},\frac{3}{4},1\} \times\{0,\frac{1}{4},\frac{2}{4},\frac{3}{4},1\}$. That means, the actual full state space is a unit square of $5 \times 5$ lattice points in which each social state can be represented by a lattice point, as shown in Fig.~\ref{fig:5x5lattice}. For example, the social state 3 denoted as $x_A$=($\frac{2}{4},\frac{1}{4}$) is a state where $\frac{2}{4}$ of subjects in the first population choose $X_{1}$ and $\frac{1}{4}$ of subjects in the second population choose $Y_{1}$. It is clear that, at a given round (time) in an experiment, the social state can be observed~\cite{Nowak2012,xu2012test,XuWang2011ICCS,Friedman2012}.

\subsection{State-dependent payoff}

There are a total of four possible strategies in the two populations. Suppose $x$ ($y$) is the proportion of the row (column) population using strategy $X_{1}$ ($Y_{1}$). It is natural that the density of strategy $X_2$ ($Y_2$) is $1-x$ ($1-y$). According to the payoff matrices of the games (see Table~\ref{fig:seltenpayoff}), the average payoff for the four strategies can be expressed as the following matrix:
  \begin{eqnarray}\label{eq:eachU}
 \left[\begin{array}{c}
U_{X_1}\\
U_{X_2}\\
U_{Y_1}\\
U_{Y_2}
\end{array}\right]
=
\left[\begin{array}{c}
a_{11}\, y + a_{12}\, \left(1 - y\right)\\
a_{21}\, y + a_{22}\, \left(1 - y\right)\\
b_{11}\, x + b_{21}\, \left(1 - x\right)\\
b_{12}\, x + b_{22}\, \left(1 - x\right)
\end{array}\right]
\end{eqnarray}
where $U_{X_{i}}$ ($U_{Y_{i}}$), $i=1,2$ is the payoff for each player in the row (column) population choosing strategy $X_{i}$ ($Y_{i}$).

Further, the average payoff for each population can be aggregated as:
  \begin{eqnarray}\label{eq:meanU}
 \left[\begin{array}{c}
\bar{U}_X \\
\bar{U}_Y
\end{array}\right]
=
\left[\begin{array}{c}
x\,U_{X_1}+\left(1 - x\right)\,U_{X_2}\\
y\,U_{Y_1}+\left(1 - y\right)\,U_{Y_2}
\end{array}\right] .
\end{eqnarray}
Therefore, if the social state at $t$ is at $(x,y)$, according to the parameters of the payoff matrix of a given game, the average payoff for each strategy and the average payoff for each population can be calculated~\cite{Sigmund1981,selten2008}.

\subsection{Eigenvalues}\label{eigenvalues}

To evaluate the cyclic social behaviours in the experimental games, we focus on the eigenvalues of the ordinary differential equations (ODE) of evolutionary dynamics. The systems are non-linear first-order differential equations. Mathematically, the eigenvalues, which are calculable, determine the dynamic properties of a system. To compare the experimental results with some earlier works~\cite{Sigmund1981,Hofbauer1996,Maynard1982evolution,Taylor1978}, we calculate the eigenvalues of the 12 games for the two dynamics.

\subsubsection{Standard replicator dynamics}
The ODE of the standard replicator dynamics is~\cite{Taylor1978,Sigmund1981}
  \begin{eqnarray}   
  \left\{
  \begin{array}{ccc}
 \dot{x} &=& ~x(U_{X_{1}}-\bar{U}_X) \\
  \dot{y} &=& y(U_{Y_{1}}~-\bar{U}_Y)
  \label{eq:replicator}
  \end{array}
  \right.
\end{eqnarray}
where ($x,y$) indicates a social state, $U$ and $\bar{U}$ have the same implications as in Eq.~\ref{eq:eachU} and Eq.~\ref{eq:meanU}, respectively. The ODE describes how the social state changes according to the payoff of the strategies. For the $2\times2$ games whose payoff matrices are shown as Fig.~\ref{fig:seltenpayoff}, according to Eq.~\ref{eq:eachU} and Eq.~\ref{eq:meanU}, the ODE above can be presented as
\begin{eqnarray}  \label{eq:replicatorA11}
\left\{
  \begin{array}{ccc}
   \dot{x} &=& - x\, \left(x - 1\right)\cdot ~~~~~~~~~~~~~~~~~~~~~~ ~~~~~~~~~~~~~~~~ \\
   && \left(a_{12} - a_{22} + a_{11}\, y - a_{12}\, y - a_{21}\, y + a_{22}\, y\right)  \\
  \dot{y} &=& - y\, \left(y - 1\right)\cdot ~~~~~~~~~~~~~~~~~~~~~~ ~~~~~~~~~~~~~~~~  \\
  &&\left(b_{21} - b_{22} + b_{11}\,  x - b_{12}\, x - b_{21}\, x + b_{22}\, x\right)
\end{array}
  \right.
\end{eqnarray}

If we let $\dot{x}=0$ and $\dot{y}=0$, then there exists a unique inner Nash equilibrium of standard replicator dynamics ($N_s$) in the rectangular coordinate system constructed by $x$ and $y$ ($0 < x, y < 1$),  which is
\begin{equation}\label{NashPoint}
      N_s =  \left(   \frac{b_{22}-b_{21} }{b_{11} - b_{12} - b_{21} + b_{22}},  \frac{ a_{22} - a_{12}}{a_{11} - a_{12} - a_{21} + a_{22}}  \right)
\end{equation}
We calculate the Nash equilibrium for the 12 games, denoted by ($x_N,y_N$), as Table~\ref{payoffmatrix12andFreq} shows. Then surrounding the $N_s$, the Jacobian is

\begin{equation}\label{JacobianSRD}
J^s = \left[\begin{array}{cc} 0 & J_{12}^{s}\\  J_{21}^{s} & 0 \end{array}\right]
\end{equation}
 in which:
\begin{equation}\label{JacobianRD12}
 J_{12}^{s} = -\frac{\left(b_{11} - b_{12}\right)\, \left(b_{21} - b_{22}\right)\, \left(a_{11} - a_{12} - a_{21} + a_{22}\right)}{{\left(b_{11} - b_{12} - b_{21} + b_{22}\right)}^2}
 \end{equation} and
\begin{equation}\label{JacobianRD21}
 J_{21}^{s} = -\frac{\left(a_{11} - a_{21}\right)\, \left(a_{12} - a_{22}\right)\, \left(b_{11} - b_{12} - b_{21} + b_{22}\right)}{{\left(a_{11} - a_{12} - a_{21} + a_{22}\right)}^2}
 \end{equation}
Then, the square of the eigenvalues   of  the Jacobian of the standard replicator dynamics, denoted as $\lambda_{s}^{2}$, is
\begin{equation}\label{eq:lambda2SRD}
 \lambda_{s}^2 =  \frac{\left(a_{11} - a_{21}\right)\, \left(a_{12} - a_{22}\right)\, \left(b_{11} - b_{12}\right)\, \left(b_{21} - b_{22}\right)}{\left(a_{11} - a_{12} - a_{21} + a_{22}\right)\, \left(b_{11} - b_{12} - b_{21} + b_{22}\right)}
\end{equation}
Plugging the payoff matrixes of the 12 games into Eq.\ref{eq:lambda2SRD}, we find that the squares of the eigenvalues are all negative,  as shown in the  $\lambda_{s}^2$ column in Table \ref{payoffmatrix12andFreq}.

\subsubsection{Adjusted replicator dynamics}
The alternative version is called adjusted replicator dynamics (also called as payoff-normalised replicator dynamics in the economics literature)~\cite{Traulsen2005,Weibull1997,Traulsen2009,Bowles2004,Sigmund2010}. Its ODE can be expressed as
\begin{eqnarray}
\left\{
  \begin{array}{ccc}
  \dot{x} &=& x(U_{X_{1}}-\bar{U}_X) \bar{U}_X^{-1} \\
  \dot{y} &=& y(U_{Y_{1}}-\bar{U}_Y)\bar{U}_Y^{-1}
  \label{eq:nprd}
   \end{array}
  \right.
\end{eqnarray}

Obviously, different to the replicator dynamics shown in Eq.~\ref{eq:replicator}, the item $\bar{U}_X^{-1}$ and $\bar{U}_Y^{-1}$ in Eq.~\ref{eq:nprd} make the payoffs normalised. Specified for $2\times2$ games whose payoff matrices are shown in  Fig.~\ref{fig:seltenpayoff}, according to Eq.~\ref{eq:eachU} and Eq.~\ref{eq:meanU}, the ODE of the adjusted replicator dynamics can be presented as
\begin{eqnarray}\label{eq:ARD}
\left\{
   \begin{array}{ccc}
  \dot{x}&=& x\left(1-x\right)\left[\left(a_{12}-a_{22}\right)+A y\right]\cdot ~~~~~~~~~~~~~~~~~~~~~ \\
 && [A xy+(a_{12}-a_{22})x+(a_{21}-a_{22})y+a_{22}]^{-1} \\
  \dot{y}&=& y\left(1-y\right)\left[\left(b_{21}-b_{22}\right)+B x\right]\cdot   ~~~~~~~~~~~~~~~~~~~~~ \\ && [B xy+(b_{21}-b_{22})y+(b_{12}-b_{22})x+b_{22}]^{-1}
     \end{array}
  \right.
\end{eqnarray}
where we have introduced the short hand
 $
A = a_{11}-a_{21}+a_{22}-a_{12}$
and
$B = b_{11}-b_{12}+b_{22}-b_{21}$.

\begin{figure*}
\centering
\resizebox{.8\textwidth}{!}{
\includegraphics{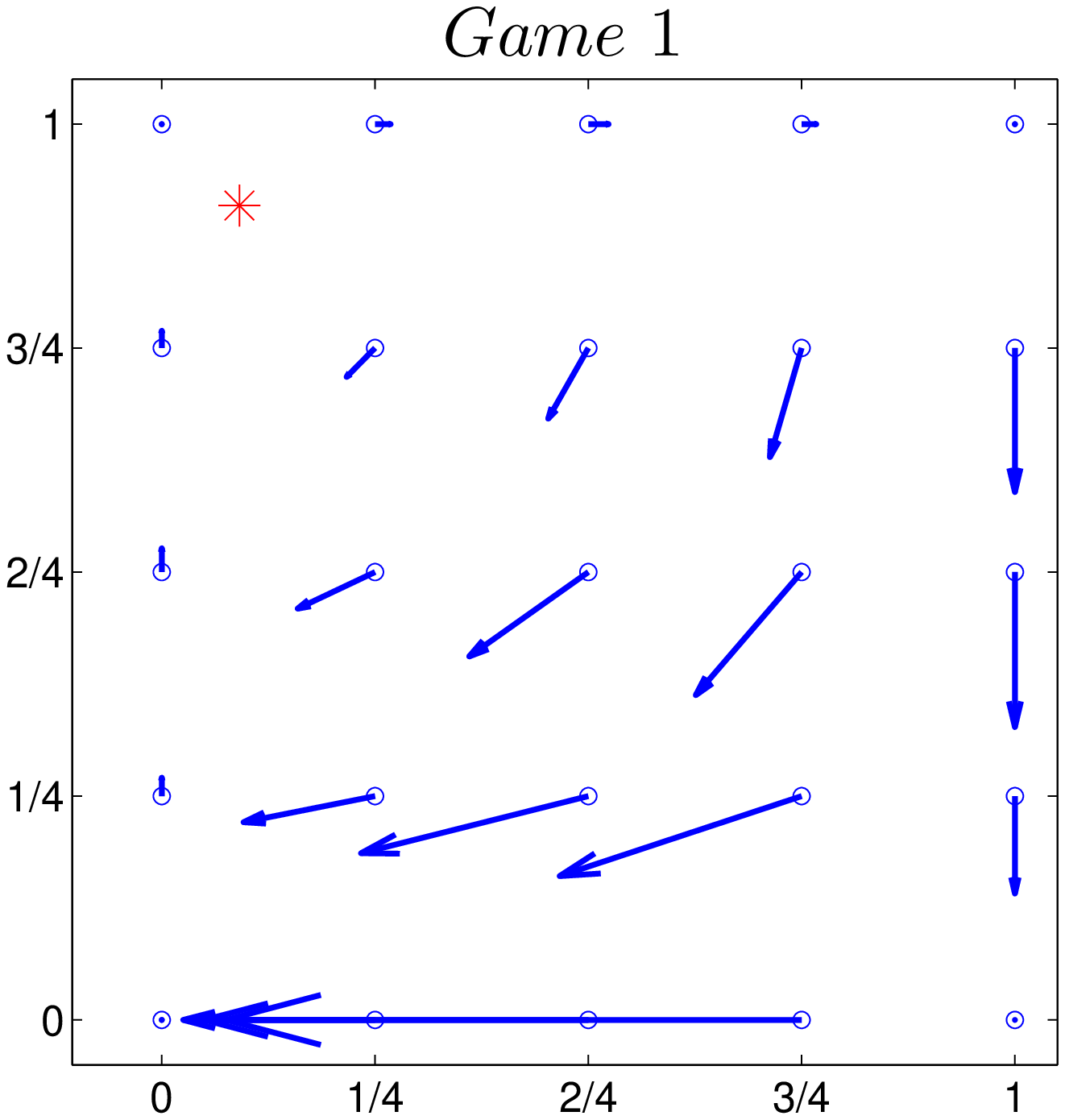}
\includegraphics{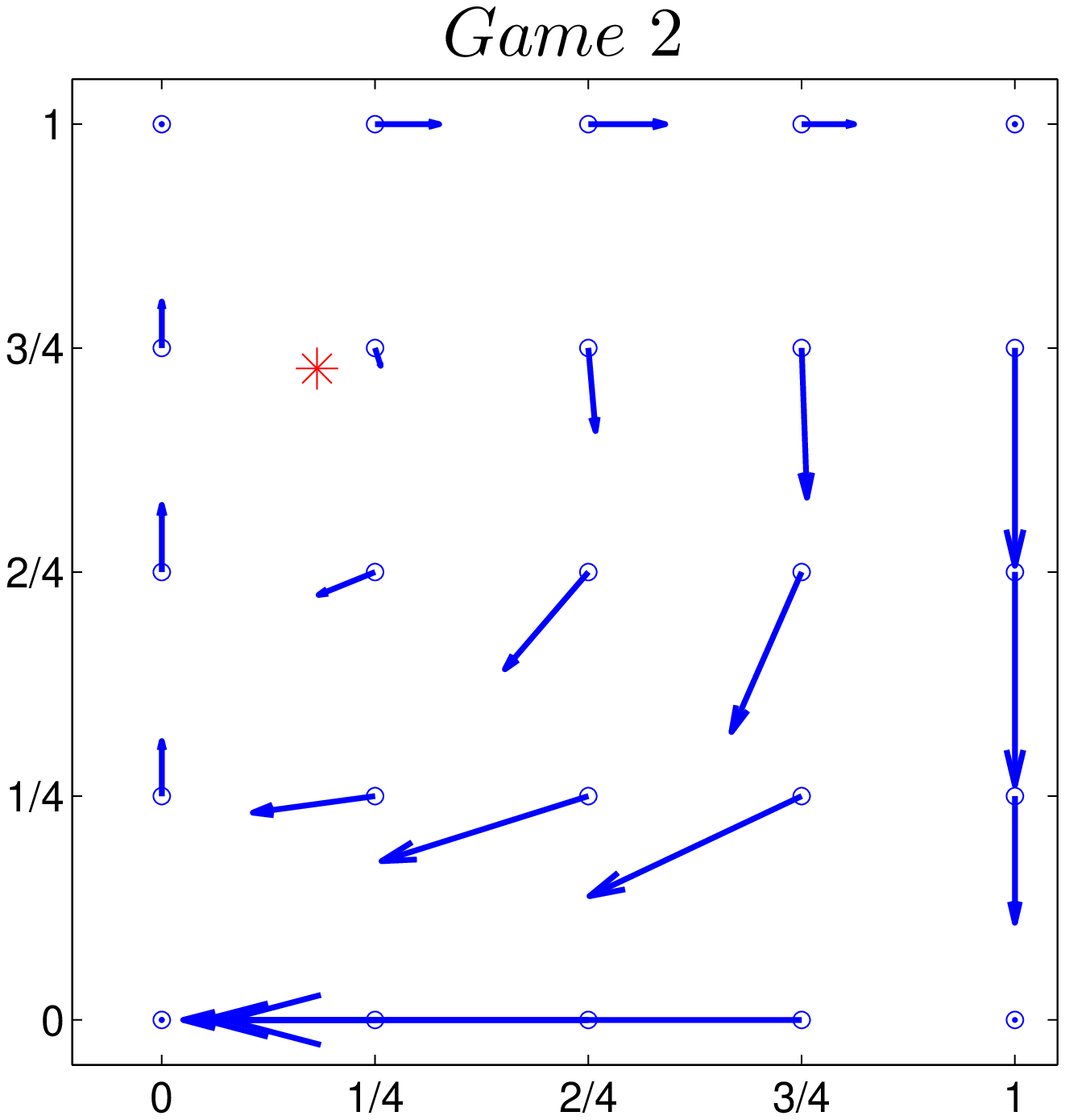}
\includegraphics{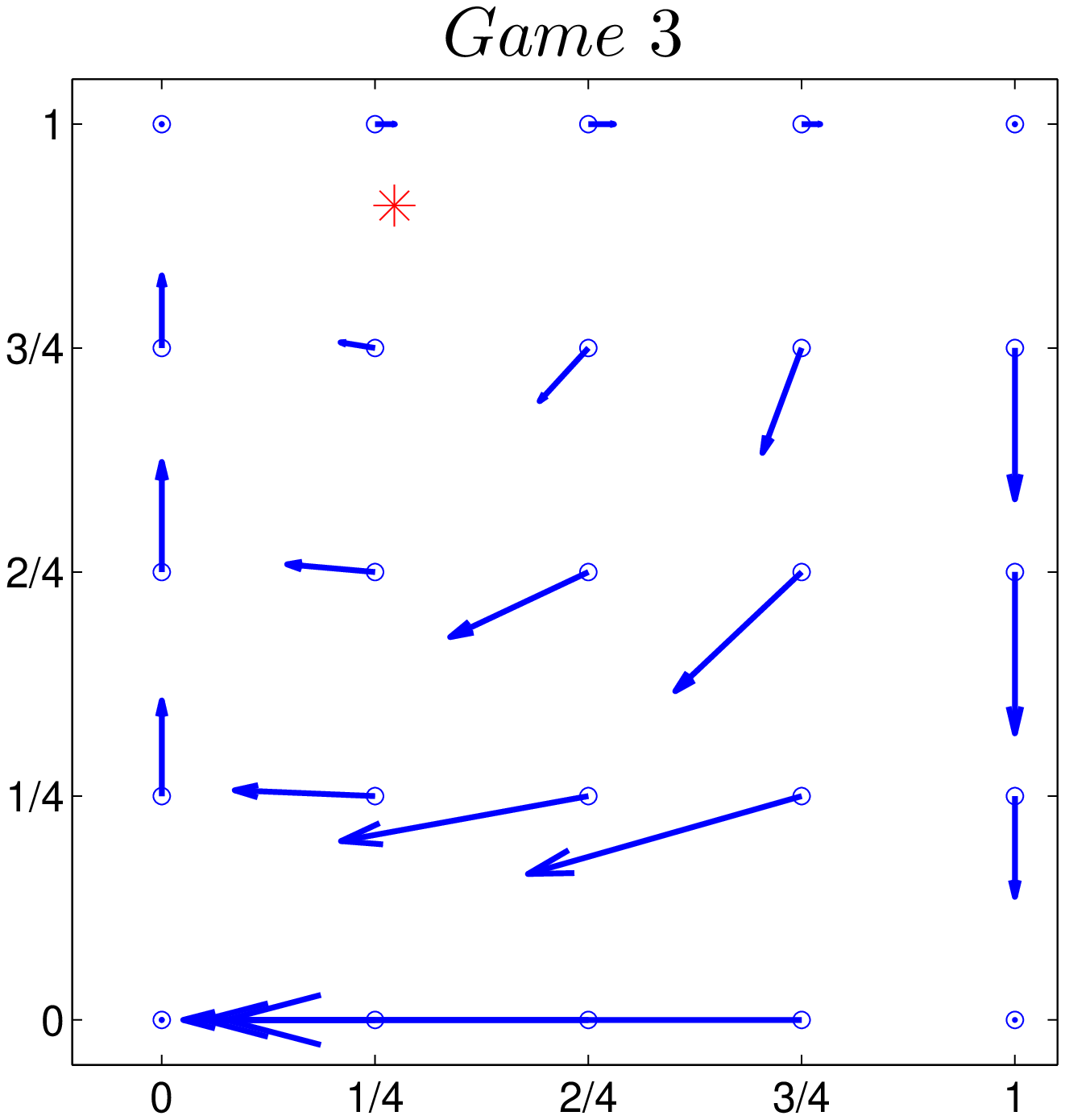}
\includegraphics{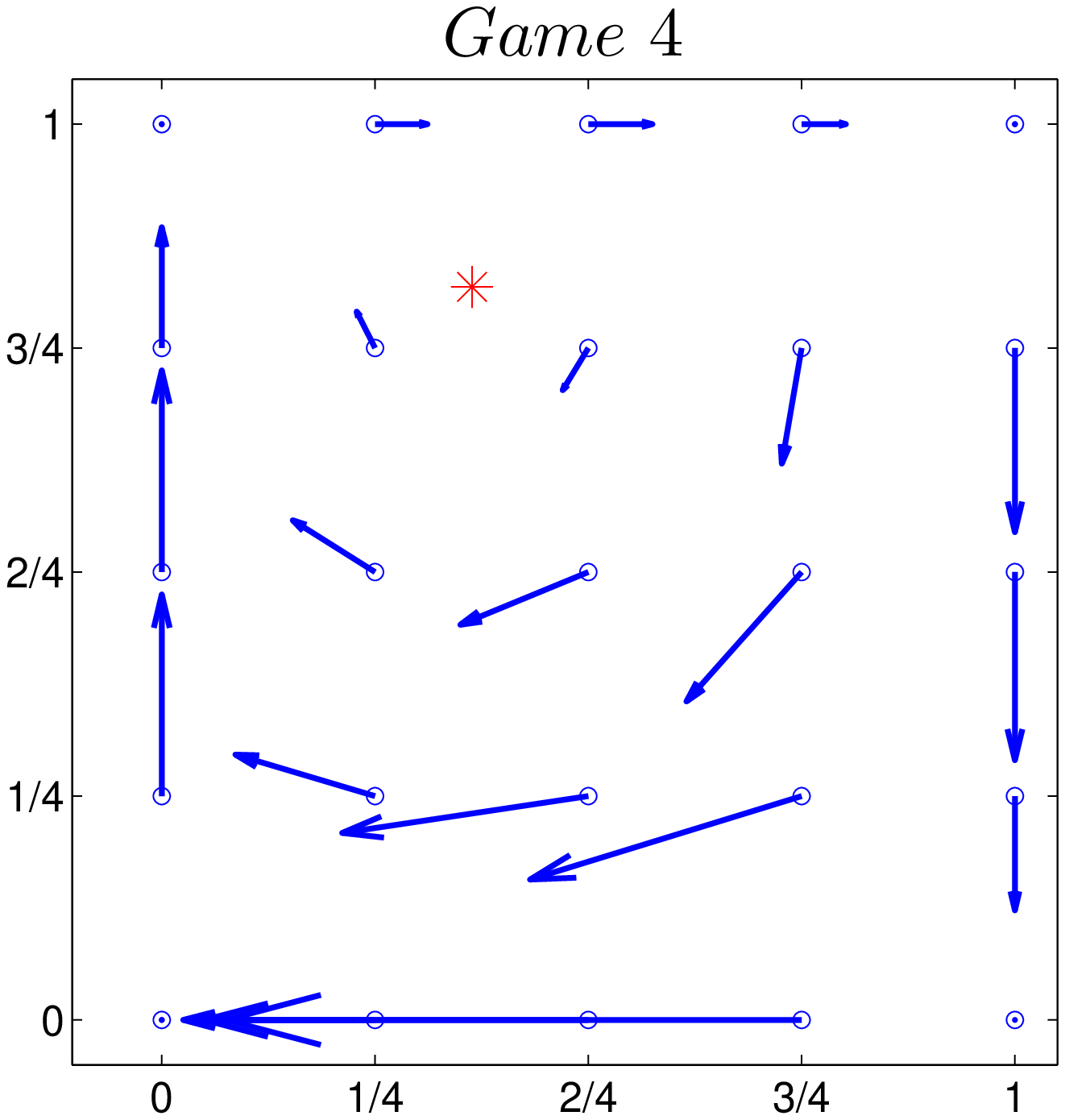}
\includegraphics{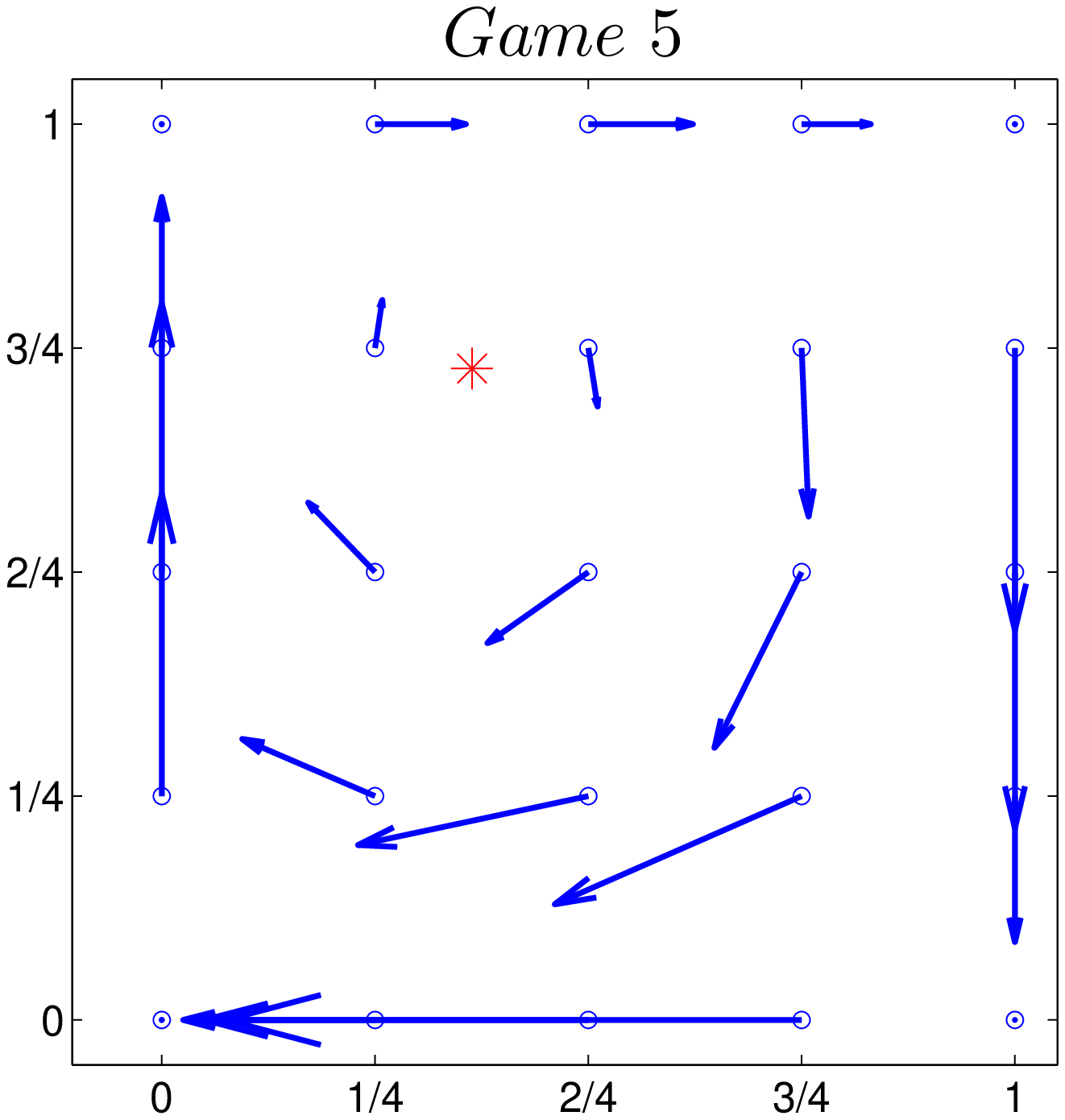}
\includegraphics{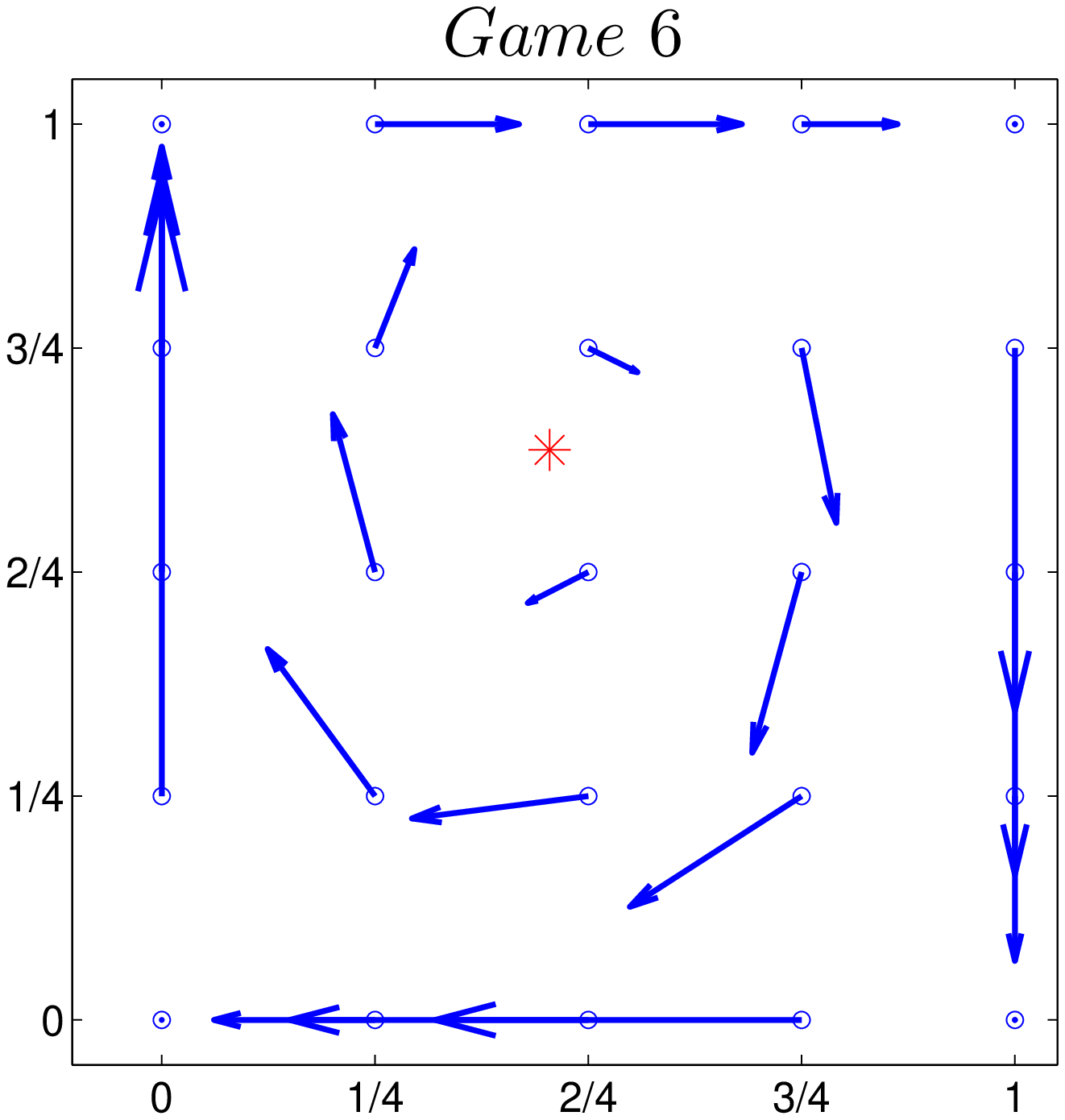}
 }
 \resizebox{0.8\textwidth}{!}{
 \includegraphics{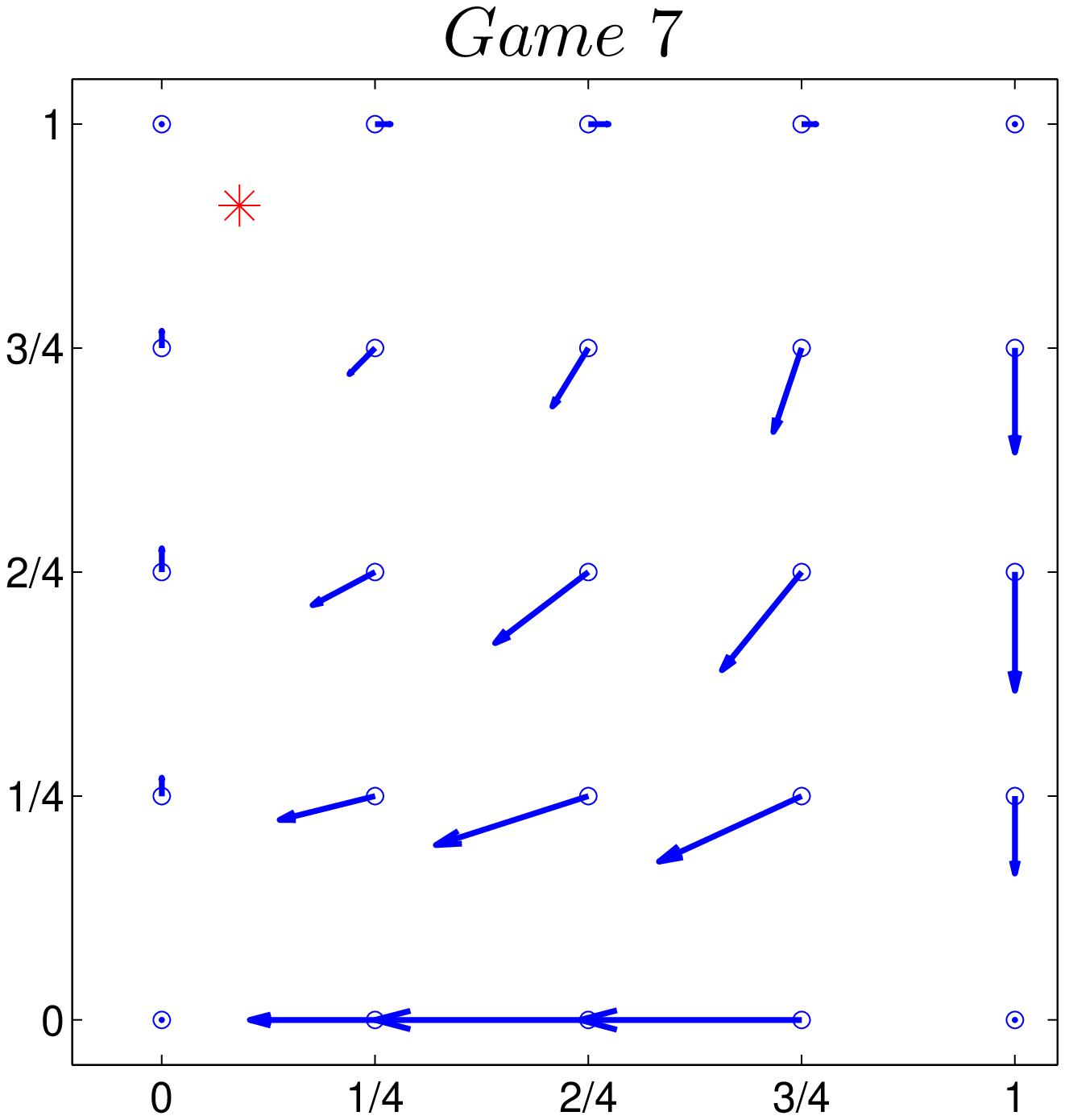}
\includegraphics{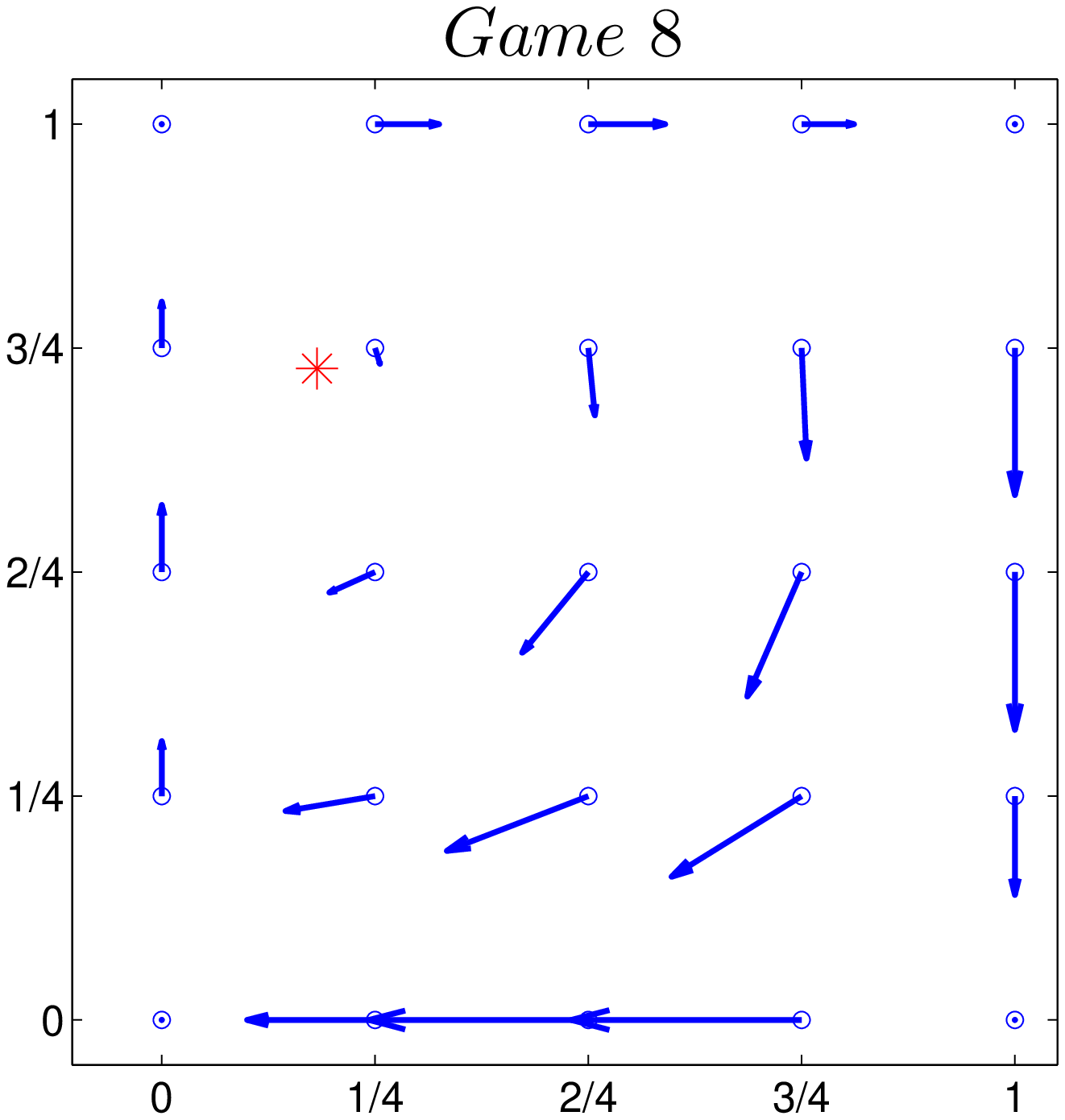}
\includegraphics{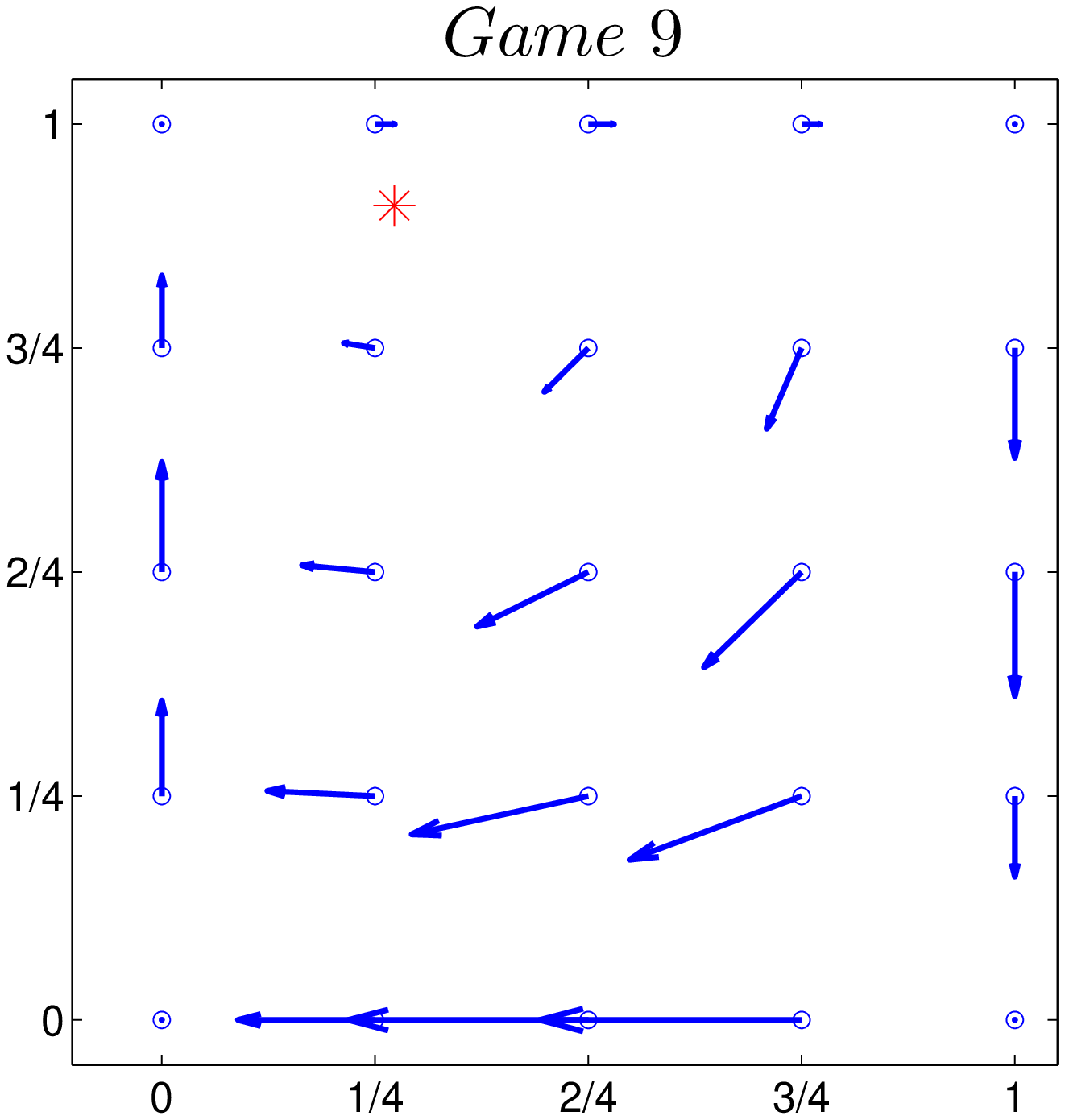}
\includegraphics{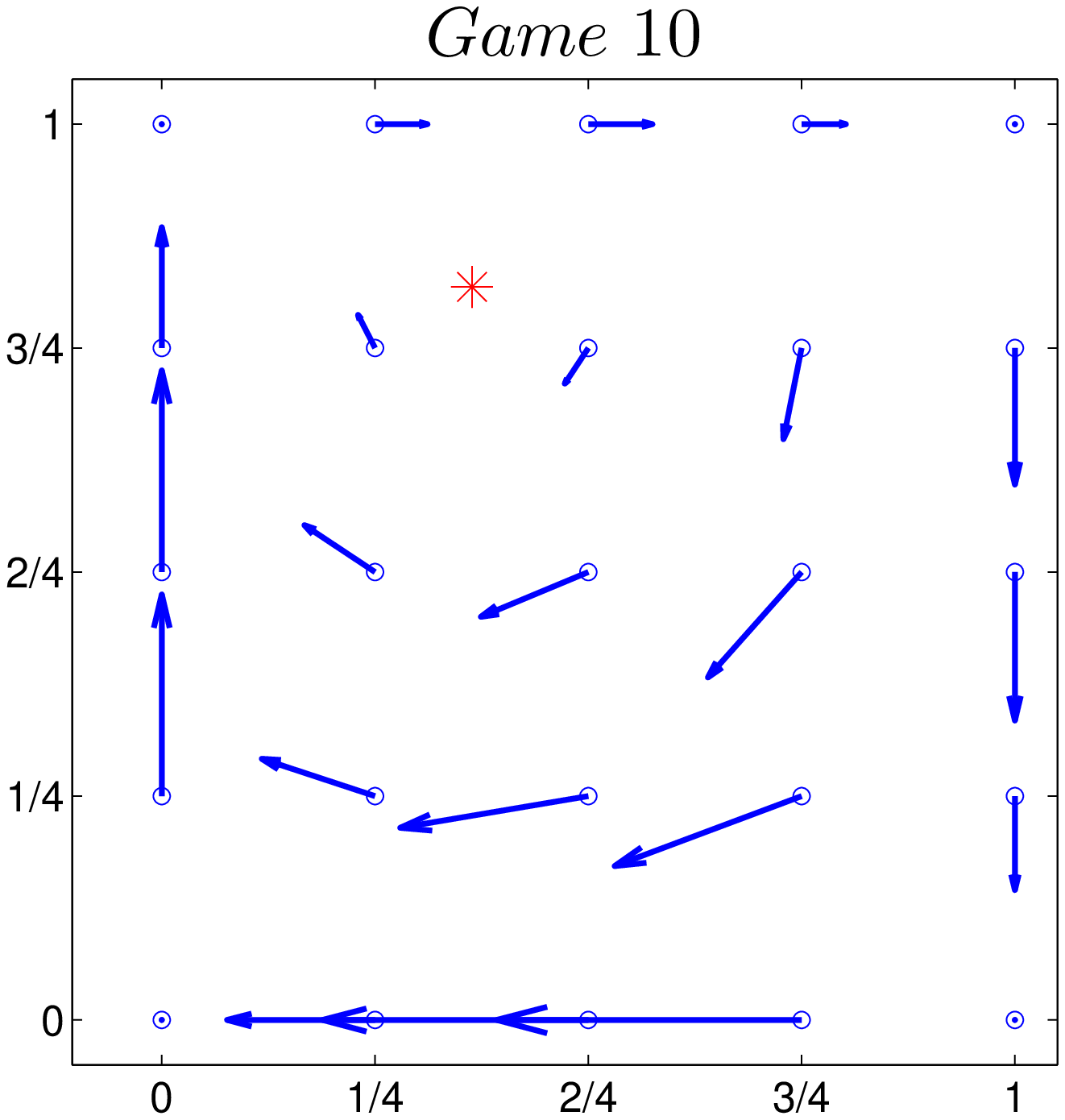}
\includegraphics{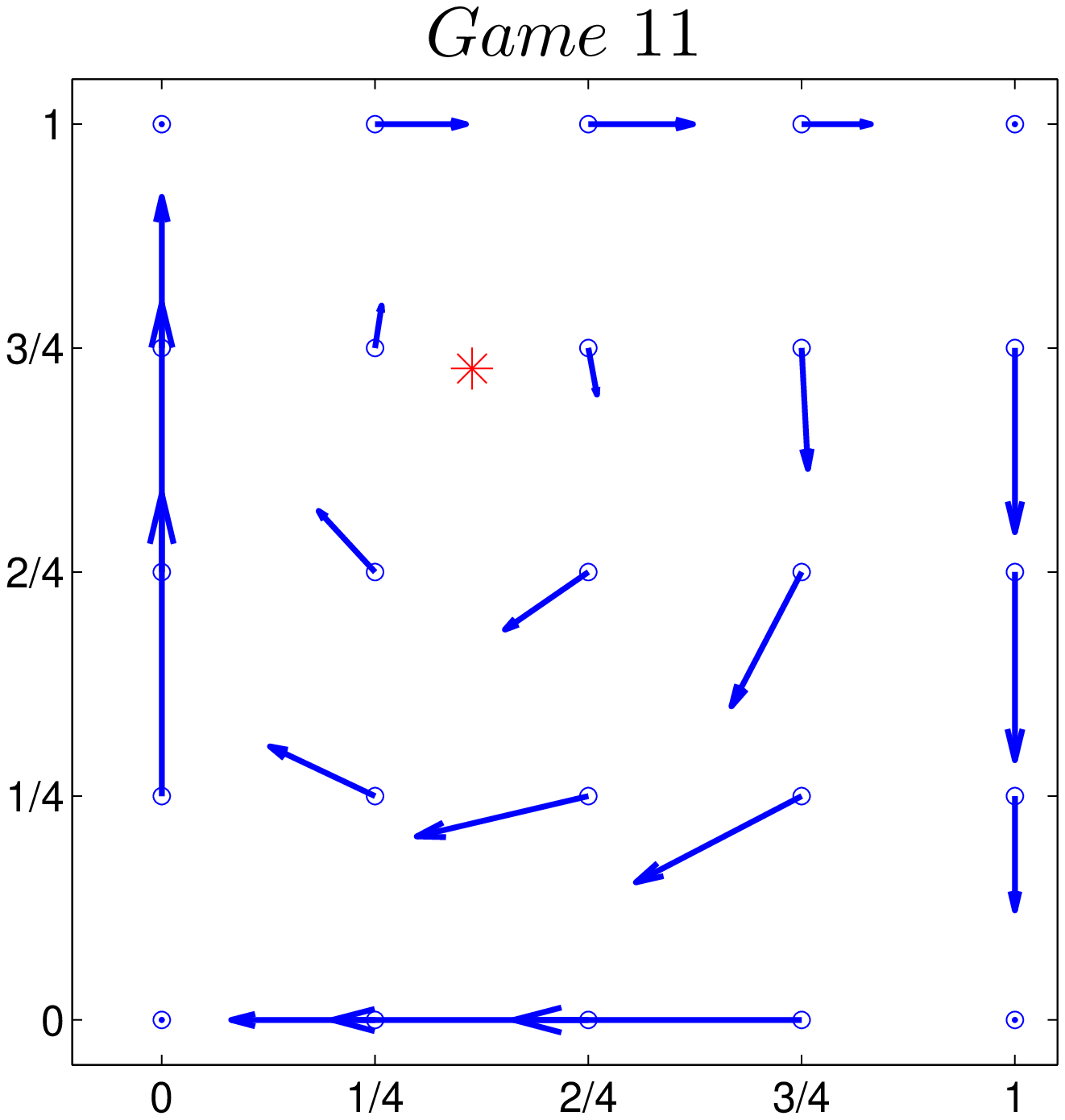}
\includegraphics{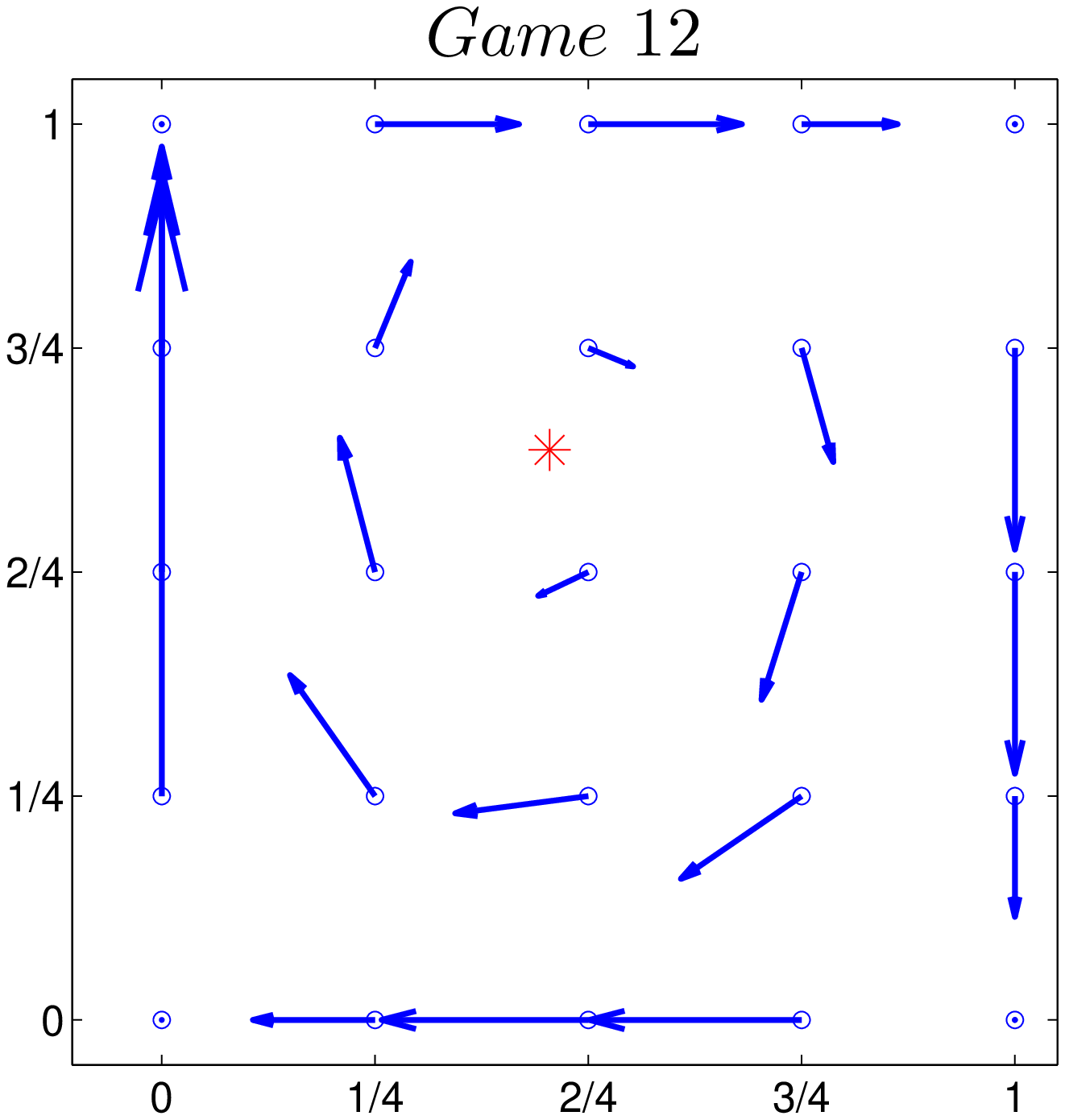}}
\caption{The expected velocity field of the adjusted replicator dynamics. The arrow  of the vector at each site represents the direction of the transit. The magnitude of the vector at each site represents the strength of the transit. At the 25 observable states, a two dimensional vector can be plotted with Eq.~\ref{eq:ARD} and the payoff matrix elements in Table~\ref{payoffmatrix12andFreq}. Surrounding the Nash equilibrium point (red cross), all of the global patterns of the 12 games are of clockwise.
\label{fig:12DynaSelten}}
\end{figure*}

For adjusted replicator dynamics, there also exists a unique inner Nash equilibrium $N_a$ in the rectangular coordinate system constructed by $x$ and $y$ ($0 < x, y < 1$),  which is
\begin{equation}\label{PQ}
      N_a=  \left(\frac{b_{21}-b_{22}}{b_{21}-b_{22}+b_{12}-b_{11}}, \frac{a_{12}-a_{22}}{a_{12}-a_{22}+a_{21}-a_{11}}\right).
\end{equation}
As $N_a$ in Eq.~\ref{PQ} is equal to $N_s$ in Eq.~\ref{NashPoint}, we have the result that the Nash equilibria of the two models are identical.

Again, surrounding the equilibrium point $N_a$, the Jacobian of  Eq.~\ref{eq:ARD} is
\begin{equation}\label{JacobianARD}
 J^a =  \left[\begin{array}{cc}
0&J_{12}^{a}\\
J_{21}^{a}&0
\end{array}\right].
\end{equation}
in which:\\
 \begin{equation}\label{J12}
 J_{12}^{a}= \frac{(b_{11}-b_{12})(b_{21}-b_{22})(a_{12}-a_{22}+a_{21}-a_{11})^{2}}{(a_{12}a_{21}-a_{11}a_{22})(b_{21}-b_{22}+b_{12}-b_{11})^{2}}
 \end{equation}
\\ and
 \begin{equation}\label{J21}
J_{21}^{a} = \frac{(a_{11}-a_{21})(a_{12}-a_{22})(b_{21}-b_{22}+b_{12}-b_{11})^{2}}{(b_{21}b_{12}-b_{11}b_{22})(a_{12}-a_{22}+a_{21}-a_{11})^{2}}.
\end{equation}
Then, the square of the eigenvalues of the Jacobian of the adjusted replicator dynamics, denoted as $\lambda_{a}^{2}$, is
\begin{equation}\label{eq:lambda2ARD}
 \lambda_{a}^{2}=\frac{  \left(a_{11} - a_{21}\right) \left(a_{12} - a_{22}\right)  \left(b_{11} - b_{12}\right)  \left(b_{21} - b_{22}\right)}{\left( a_{12}\, a_{21} - a_{11}\, a_{22} \right)  \left( b_{21}\, b_{12} - b_{11}\, b_{22}\right)}
\end{equation}
For the 12 games, the results of $\lambda_{a}^2$ are shown in Table~\ref{payoffmatrix12andFreq}.

One point worth emphasising is that since all the $\lambda^2$  are negative, the eigenvalues should be purely imaginary. So, the solution of the dynamics (Eq.~\ref{eq:replicatorA11} and Eq.~\ref{eq:ARD} near Nash equilibrium) is cyclic and equivalent to simple harmonic motion (SHM). The following part explains this point in more detail.

We use $(\triangle x, \triangle y)$ to present the deviation from the Nash Equilibrium of $(x, y)$. Referring to the Jacobian (Eq.~\ref{JacobianARD} for adjusted replicator dynamics),  we have
\begin{equation}\label{Eq2DSHMJ}
  \left[\begin{array}{c}\dot{\triangle x} \\ \dot{\triangle y}\end{array}\right]   =    \left[\begin{array}{cc}
0&J_{12}^{a}\\
 J_{21}^{a}&0
\end{array}\right]  \left[\begin{array}{c} \triangle x  \\  \triangle y \end{array}\right].  \\
\end{equation}
From time derivation of this equation, we have
\begin{equation}\label{Eq2DSHMJa}
\left[\begin{array}{c}\ddot{\triangle x} \\ \ddot{\triangle y}\end{array}\right]  =   \left[\begin{array}{cc}
J_{12}^{a}J_{21}^{a}&0\\
0&J_{12}^{a}J_{21}^{a}
\end{array}\right] \left[\begin{array}{c} \triangle x  \\  \triangle y \end{array}\right].  \\
\end{equation}

Mathematically, this dynamics is equivalent to \emph{asymmetric games with cyclic dynamics} (see Eq.~J.~3 in Appendixes J in  textbook~\cite{Maynard1982evolution}) and the solutions of this dynamics are closed cyclical orbits.

\subsection{Theoretical expectations}\label{secTheoVal}

The theoretical expectations of directions, frequencies and changes in radius of the cycles can be obtained from the eigenvalues (in Eq.~\ref{eq:lambda2SRD} and Eq.~\ref{eq:lambda2ARD}) as well as the Jacobian (in Eq.~\ref{JacobianSRD} and Eq.~\ref{JacobianARD}).

\subsubsection{Directions of cycles}
In  state space, the direction of cycles could be clockwise or counter-clockwise. We use two methods to evaluate it.

One way is to plot the velocity field on the basis of Eq.~\ref{eq:replicatorA11} and Eq.~\ref{eq:ARD}. As shown in Fig.~\ref{fig:12DynaSelten}, all directions of the velocity fields are clockwise. Another way is using the Jacobian described~\cite{Sigmund1981} before. Calculating the Jacobian (Eq.~\ref{JacobianSRD} or Eq.~\ref{JacobianARD}) explicitly, we can get the changing direction at each given state around $N_s$ (or $N_a$). For the two dynamics, as $J_{12} > 0$ and $J_{21} < 0$ hold for all the 12 games, we have all the directions in a clockwise manner.

At the same time, it also explains the existence and persistence of the cycles in the same ways.

\subsubsection{Frequencies of cycles}
As mentioned above,  the eigenvalue $\lambda$ (denoted as $\lambda=\alpha \pm i\beta$) of the 12 games  is purely imaginary, i.e. the real part ($\alpha$) is zero  for all $\lambda$. Mathematically, the motion is equivalent to SHM, in which the value of $\beta$ refers to  the \emph{angular frequency} (also called angular speed, see page 199 in~\cite{holzner2011physics}). Since there are 2$\pi$   radians in one cycle, in order to complete the cycle $2\pi/\beta$ units of time are needed. So the theoretical periodic frequency ($f_{theo}$) is explicitly given as:
\begin{equation}\label{eq:frequency}
 f_{theo}=\frac{\beta}{2\pi}
\end{equation}
For the 12 games, the values of $f_{theo}$ for standard and adjusted replicator dynamics (denoted as $f_s$ and $f_a$, respectively) are different (shown in Table~\ref{payoffmatrix12andFreq}). They will serve as the theoretical expectations (in Table~\ref{tab:OLRresults} and Fig.~\ref{fig:scatter}) to be compared with observed frequencies.

\subsubsection{Radius of cycles}
As  noted, for all of the 12 games and both of the models, the real parts of the eigenvalues ($\alpha$) are equal to 0. As mentioned above, motions of these game systems are mathematically equivalent to simple harmonic motion. Since $\alpha=0$, the radius of cycles does not change with time.

In the 200-round experiments, the testable hypothesis is: the cycle radius of the first 100 rounds equals that of the second 100 rounds.

\section{Evaluation of the two replicator dynamics}\label{ModelEval} 

This part is the supporting material for Table~\ref{tab:OLRresults}. The statistic method and the comparison results are explained as follows.

\subsection{Statistic method}
\begin{figure}
\centering
\resizebox{0.40\textwidth}{!}{
\includegraphics {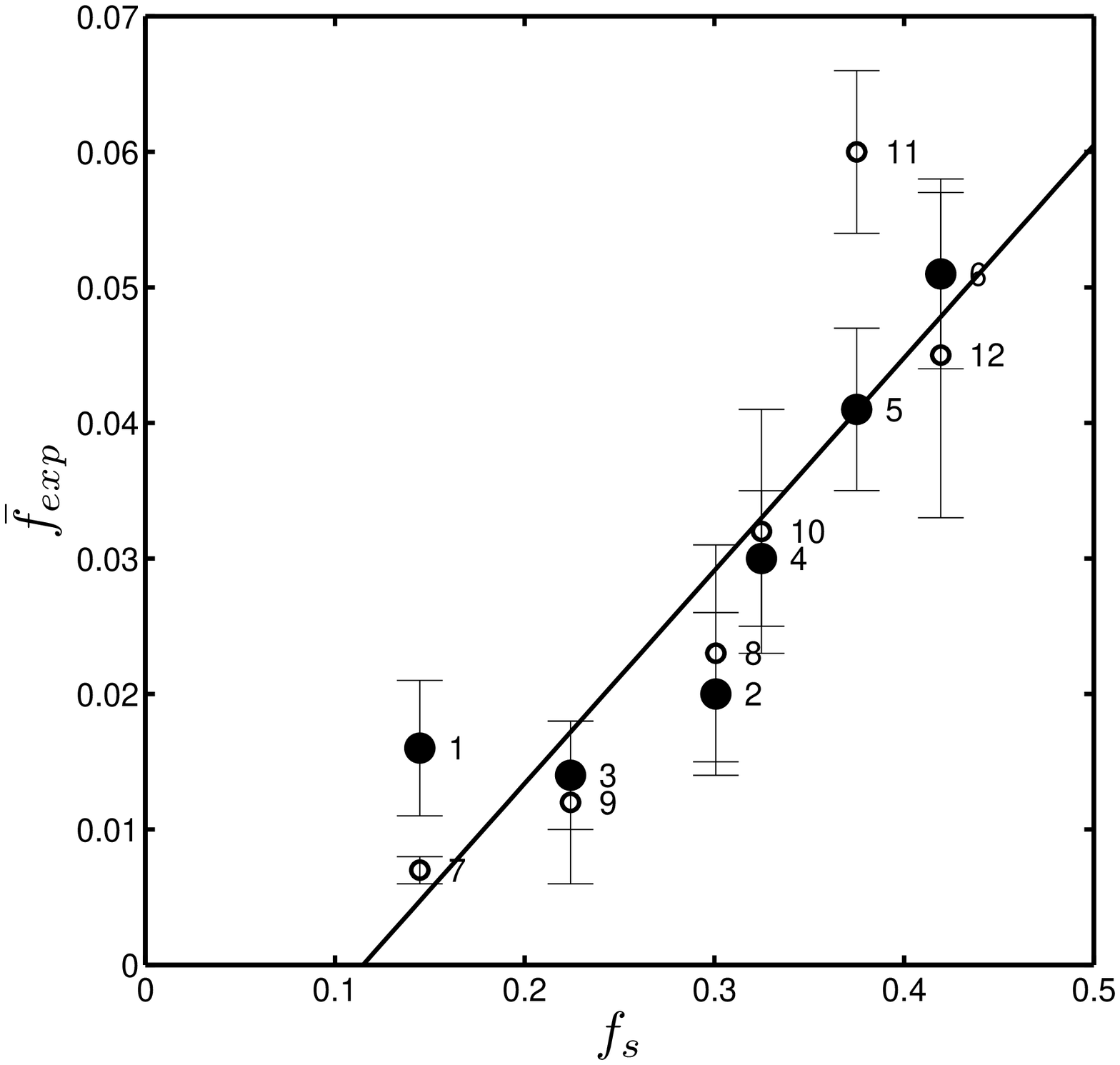}}   
\resizebox{0.40\textwidth}{!}{
\includegraphics {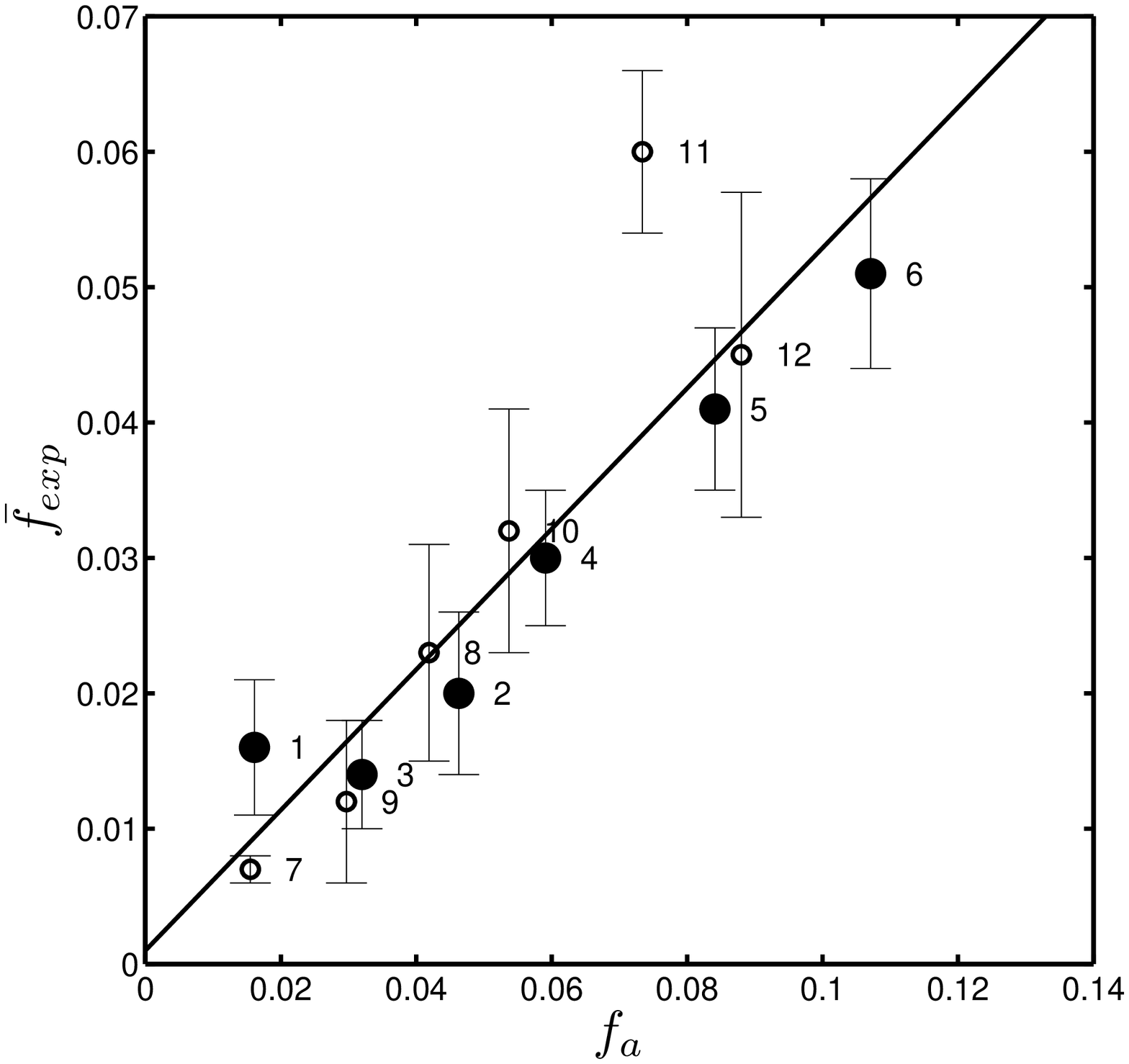}
}
\caption{The relationship between empirical frequencies and theoretical frequencies. The $x$-axis ($y$-axis) indicates the theoretical frequency $f_{theo}$ (the experimental frequency $\bar{f}_{exp}$). The $f_s$ in the panel above  comes from standard replicator dynamics,  and the $f_a$ in the panel below is from  adjusted replicator dynamics.  The black dots indicate the mean of each 12-group game (games 1 to 6) and the black holes of each 6-group game (games 7 to 12). The bars indicate the s.e.m (standard error of the mean) and the numbers correspond to the game ID. The slope lines are plotted according to the results of OLR. The adjusted replicator dynamics perform significantly better than the standard replicator dynamics.
\label{fig:scatter}}
\end{figure}

Referring to the theoretical frequency ($f_{s}$ and $f_{a}$ in Table~\ref{payoffmatrix12andFreq}) and the experimental frequencies ($\bar{f}_{exp}$ in Table~\ref{tab:theta}), the OLR results are shown in Table~\ref{tab:OLRresults}. The relationship between the theoretical frequencies ($f_{theo}$) and experimental frequencies ($\bar{f}_{exp}$) of the 12 games can also be demonstrated as in Fig.~\ref{fig:scatter}. The statistical testing and analysis are explained as follows.

(1) For the comparison with $f_{s}$ from the standard replicate dynamics, statistical results show that the coefficient of the linear regression over the 12 paired samples is $0.157\pm0.026$. It indicates that the positive correlation between theoretical frequencies and experimental frequencies is strongly significant (at $t>5$ level).
In addition, the constant of the regression is $-0.018\pm0.008$ ($p=0.056>0.05$) and its $95\%$ C.I. [-0.036  0.shi001] includes zero. This implies that there is no systematic deviation.

(2) Comparing with $f_{a}$ from the adjusted replicate dynamics, statistical results show that the coefficient of the linear regression over the 12 paired samples is $0.519\pm0.081$. This indicates that the positive correlation between theoretical frequencies and experimental frequencies is strongly significant (at $t=6.41>5$ level).
In addition, the constant of the regression is $0.001\pm0.005$ ($p=0.805$, two-tailed) and its $95\%$ C.I. [-0.010 0.012] includes 0.000. This implies that there is no systematic deviation here either.

\subsection{Comparison results}
The comparison results are as follows.

(1) According to the $p$-value  of the coefficient of the linear regression, both of the models' expectations are positively related to experimental observations, and can infer the difference of the observed frequencies well. Since the constant is closer to 0 and the $t$-value (see Table~\ref{tab:OLRresults}) is larger, the adjusted replicator dynamics  therefore \emph{performs better}.

(2) However, \emph{neither} of the two models can capture the observed frequencies quantitatively and exactly, because the values of the coefficients of the linear regression for both  the standard replicator dynamics  ($0.157\pm0.026$)   and  the adjusted replicator dynamics  ($0.519\pm0.081$) are significantly different from 1.

\bibliographystyle{epjc}

%
%
%

 \end{document}